\documentclass[preprint,12pt]{aastex}
\usepackage{epsf}
\newcommand{\simgt}{\lower.5ex\hbox{$\; \buildrel > \over \sim \;$}}
\newcommand{\simlt}{\lower.5ex\hbox{$\; \buildrel < \over \sim \;$}}
\newcommand{\calF}{{\mbox{\boldmath${\cal F}$}}}
\newcommand{\bfv}{{\mbox{\boldmath$v$}}}
\newcommand{\bfx}{{\mbox{\boldmath$x$}}}
\newcommand{\bfp}{{\mbox{\boldmath$p$}}}
\newcommand{\bfq}{{\mbox{\boldmath$q$}}}
\newcommand{\rangleL}{\rangle_{\scriptscriptstyle L}}
\newcommand{\rangleE}{\rangle_{\scriptscriptstyle E}}
\newcommand{\probL}{P_{\scriptscriptstyle L}}
\newcommand{\probE}{P_{\scriptscriptstyle E}}
\newcommand{\probI}{P_{\scriptscriptstyle I}}
\slugcomment{RESCEU-18/02\quad UTAP-429/2002}
\shortauthors{Ohta,Kayo\& Taruya}
\shorttitle{Cosmological Density Distribution From Local Collapse Model}
\begin{document}
\title{Evolution of Cosmological Density Distribution Function from the
Local Collapse Model}
%
%
\author{Yasuhiro Ohta, Issha Kayo and Atsushi Taruya}
\affil{Department of Physics and Research Center for the Early 
Universe(RESCEU), 
\\ School of Science, University of Tokyo, Tokyo 113, Japan.}
\email{ohta@utap.phys.s.u-tokyo.ac.jp, kayo@utap.phys.s.u-tokyo.ac.jp,
ataruya@utap.phys.s.u-tokyo.ac.jp}
%
%
%
%
\begin{abstract}
We present a general framework to treat the evolution of one-point
 probability distribution function (PDF) for cosmic density $\delta$ and
 velocity-divergence fields $\theta$.
In particular, we derive an evolution equation for the one-point PDFs
 and consider the stochastic  nature associated with these quantities.
Under the local approximation that the evolution of cosmic fluid fields
 can be characterized by the Lagrangian local dynamics with finite
 degrees of freedom, evolution equation for PDFs becomes a closed form
 and consistent formal solutions are constructed. 
Adopting this local approximation, we explicitly evaluate the one-point
 PDFs $P(\delta)$ and $P(\theta)$ from the spherical and the ellipsoidal
 collapse models as the representative Lagrangian local dynamics. 
In a Gaussian initial condition, while the local density PDF 
from the ellipsoidal model almost coincides
 with the that of the spherical model, 
differences between spherical and ellipsoidal 
collapse model are found in the velocity-divergence PDF. 
These behaviors have also been confirmed from the perturbative analysis
 of higher order moments. 
Importantly, the joint PDF of local density, $P(\delta,t;\delta',t')$,
 evaluated at the same Lagrangian position but at the different times
 $t$ and $t'$ from the ellipsoidal collapse model exhibits a large
 amount of scatter. 
The mean relation between $\delta$ and $\delta'$ does fail to match the
 one-to-one mapping obtained from spherical collapse model. 
Moreover, the joint PDF $P(\delta;\theta)$ from the ellipsoidal collapse
 model shows a similar stochastic feature, both of which are indeed
 consistent with the recent result from N-body simulations. 
Hence, the local approximation with ellipsoidal collapse model provides
 a simple but a more physical model than the spherical collapse model of
 cosmological PDFs, consistent with the leading-order results of exact
 perturbation theory. 
\end{abstract}
\keywords{cosmology: theory - galaxies: clustering - galaxies: dark matter 
- large-scale structure of universe - methods: analytical}
%
%
%
%
%
%
%
%
%
\section{INTRODUCTION}
\label{sec: intro}

The large-scale structure of the universe is thought to be developed by
the gravitational instability from the small Gaussian density
fluctuations. 
In a universe dominated by dark matter, the evolution of mass
distribution is entirely governed by gravitational dynamics. 
While luminous objects such as the galaxies and the clusters are
subsequently formed by complicated processes including gas dynamics and
radiative process, the dark matter distribution is the most fundamental
product in the hierarchical clustering of the cosmic structure
formation. 
In particular, the clustering feature of dark matter distribution is
directly observed via weak gravitational lensing effect
(\citealt{BS2001} for review and references there in). 
Thus, the statistical study of the large-scale mass distribution
provides a useful cosmological tool in probing the formation mechanism
of dark matter halos, as well as the observed luminous distribution.

In principle, all the statistical information of dark matter
distribution is characterized by the probability distribution functions
(PDFs) for mass density fluctuation and velocity fields, $\delta$ and
$\bfv$. 
Among these, the one-point PDF for density field, $P(\delta)$, is one of
the most fundamental statistical quantities and because of its
simplicity, there has been numerous theoretical studies on its evolution
from a Gaussian initial condition. 
From the analytical study of one-point PDFs, \citet{Ko1994} first 
calculated the PDF using the Zel'dovich approximation. 
For a perturbative construction of one-point PDF, \citet{BK1995} and
\citet{J1995} introduced the Edgeworth expansion to derive the analytic
formula for PDFs. 
On the other hand, based on the tree-level approximation, \citet{B1992b}
constructed the PDF from the generating function of the vertex
function. 
Interestingly, the vertex function in the tree-level approximation is
obtained as an exact solution of the spherical collapse model. 
The effect of smoothing has been later incorporated into his calculation
and the analytic PDF was compared with N-body simulations
\citep{B1994b}.
Following these results, \citet{FG1998a} proposed to use the spherical
collapse model for the prediction of higher-order moments beyond the
tree-level approximation. 
The most remarkable point in their approximation is that the
leading-order prediction exactly matches the one obtained from the
rigorous perturbation theory and the correction for next-to-leading
order is easily computed by solving the simple spherical collapse
dynamics. 
Further, the spherical collapse approximation is recently extended to
the prediction of one-point PDF (\citealt{SG2001}, see also
\citealt{PS1997}). 
The approximation has been tested against N-body simulations and a good
agreement was found even in a non-linear regime of the density
perturbation.

On the other hand, from the numerical study, \citet*{Ka2001}
recently showed that the log-normal PDF quantitatively approximates the
one-point PDF $P(\delta)$ in the N-body simulations emerging from the
Gaussian initial condition, irrespective of the shape of initial power
spectra. 
The log-normal PDF has been long known as an empirical model
characterizing the N-body simulations
\citep[e.g.,][]{Ko1994,BK1995,TW2000} or the observed galaxy
distribution \citep[e.g,][]{Hamilton1985,BSDFYH1993,Ko1994}. 
Recently, a good agreement with the log-normal model has been 
reported in the numerical study of weak lensing statistics 
\citep{TTHKF2002}, and an attempt to clarify the origin of the 
log-normal behavior has also made \citep*{THK2002}. 
Mathematically, the log-normal PDF is obtained from a one-to-one local
mapping between the Gaussian and the non-linear density fields. 
Thus, the successful fit to the N-body simulation might be interpreted
to imply that the evolution of local density can be well-approximated by
the one-to-one local mapping. 
Indeed, the analytic PDF for the spherical collapse approximation can
also be expressed as a one-to-one local mapping via the spherical
collapse model.

The above naive picture has been critically examined by \citet*{Ka2001}
evaluating the joint PDF $P(\delta_1, t_1;\,\delta_2, t_2)$, i.e., joint
probability of the local density fields $\delta_1$ and $\delta_2$ at the
same comoving position but at the different times $t_1$ and $t_2$. 
It has been found that a large amount of scatter in the relation between
$\delta_1$ and $\delta_2$ shows up and their mean relation significantly
deviates from the prediction from the one-to-one log-normal mapping. 
Although this might not be surprising, the good agreement between the
log-normal and the simulated PDFs becomes more contrived and
difficult to be explained in a straight forward manner.

Definitely, the failure of the one-to-one local mapping somehow reflects
the non-locality of the gravitational dynamics. 
That is, the evolution of local density cannot be determined by the
initial value of the local density only. 
Rather, the local density can be expressed as multi-variate functions of
local density and other local quantities such as velocity, gradient of
local density, velocity-divergence and so on. 
Furthermore, initial values of these local quantities are randomly
distributed over the three-dimensional space. 
As a consequence, even if the dynamics is deterministic, the relation 
between the evolved and the initial local density inevitably becomes
stochastic. 
In a sense, the situation is quite similar to the non-linear stochastic
biasing of galaxies, i.e., the statistical uncertainty between galaxies
and dark matter arising from the hidden variables
\citep*[e.g.,][]{DL1999,TKS1999,TS2000}. 
Then, taking account of this stochastic nature, the crucial issue is to
construct a simple but physically plausible model of one-point PDF, at
least, consistent with the N-body simulations in a qualitative manner. 
Further, the influence of stochasticity on the evolution of local 
quantities should be investigated.

The purpose of this paper is to address these issues starting from a 
general theory of evolution of one-point PDF. 
In particular, we derive an evolution equation for the density and the
velocity-divergence PDFs and consider how the stochastic nature of the
local density field emerges. 
Due to the incompleteness of the equations, any theoretical approach
using the evolution equations for PDFs generally becomes intractable. 
However, under the {\it local approximation} that the evolution of
density field (or velocity-divergence) is entirely determined by the
local dynamics with finite degrees of freedom, we explicitly show that
the analytical solution for the evolution equations is consistently
constructed. 
Based on this approximation, the one-point PDFs for the density and the
velocity-divergence are computed from the ellipsoidal collapse model, as
a natural extension of the one-to-one mapping of the spherical collapse
approximation. 
Further, the stochastic nature arising from the multi-variate function
of local quantities is explicitly shown evaluating the joint PDFs of the
local density and/or the velocity-divergence. 
The influence of this effect is discussed in detail comparing with the
spherical collapse approximation.

This paper is organized as follows. 
In section \ref{sec: evolve_PDF}, we consider a general framework to
treat the evolution of one-point PDFs and derive the evolution equations
for the Eulerian and the Lagrangian PDFs 
(Sec.\ref{subsec: derivation_Lag} and \ref{subsec: derivation_Eul}). 
Then, adopting the local approximation, consistent solutions for these
equations are obtained (Sec.\ref{subsec: LocalApprox}). 
Further, the stochastic nature of the evolution of PDFs is quantified
evaluating joint PDFs (Sec.\ref{subsec: Joint PDF}). 
As an application of these general considerations, in section 
\ref{sec: demonstration}, we give an explicit evaluation of one-point
PDFs adopting the spherical and the ellipsoidal collapse models as
representative Lagrangian local dynamics. 
The qualitative features of the results are compared with the
perturbative analysis presented in appendix \ref{appen: perturbation} or
the previous N-body study. 
Finally, section \ref{sec: conclusion} is devoted to the conclusion 
and the discussion. 
%
%
%
%
%
%
%
%
%
%
\section{EVOLUTION EQUATION FOR PROBABILITY DISTRIBUTION FUNCTION}
\label{sec: evolve_PDF}
%
%
%
%
%
%
\subsection{Preliminaries}
\label{subsec: preliminary}
%
%
%
%
%
%
Throughout the paper, we treat dark matter as a pressure-less and
non-relativistic fluid. 
Assuming the homogeneous and isotropic background universe, the density
field $\delta(\bfx,t)$, the peculiar velocity field $\bfv(\bfx,t)$ and
the gravitational potential $\phi$ for the fluid follow the equation of
continuity, the Euler equation and the Poisson equation as follows:
\begin{equation}
 \frac{\partial \delta}{\partial t}+
\frac{1}{a}\nabla\cdot\left\{(1+\delta)\bfv\right\}=0,
\label{continuity}
\end{equation}
\begin{equation}
 \frac{\partial\bfv}{\partial t}+\frac{1}{a} (\bfv\cdot\nabla) \bfv+H\bfv 
 = -\frac{1}{a}\nabla\phi,
\label{Dynamic}
\end{equation}
\begin{equation}
 \nabla^2\phi = 4\pi G\,\rho_{\rm m}\,a^2\delta,
\label{Poisson}
\end{equation}
where $a$ is the scale factor of the universe, $H(\equiv\dot{a}/a)$
denotes the expansion rate and $\rho_{\rm m}$ is the background mass
density.

While we are mainly interested in characterizing the large-scale
structure on the basis of  statistical properties of the density field
$\delta(\bfx,t)$ and the velocity field $\bfv(\bfx,t)$ as described in
section \ref{sec: intro}, the dynamical evolution of such quantities is
not solely determined locally. 
Hence, we must introduce the other local quantities characterizing the
non-local properties of gravitational evolution, e.g., the gradient of
local density $\nabla\delta$, the velocity tensor 
$\partial_i v/\partial x_j$, and so on.  
Let us denote these variables by
\begin{equation}
\calF\equiv\left(\delta(\bfx,t),\bfv(\bfx,t),\nabla\delta(\bfx,t),
\frac{\partial v_i}{\partial x_j}(\bfx,t),\cdots\right),
\end{equation}
and define the joint PDF, $P(\calF;t)$, which gives a probability
density that the quantity $\calF$ takes some values of
$(\delta,\bfv,\nabla\delta,\cdots)$ at the time $t$. 
In terms of this, the one-point PDF of the density fluctuations
$P(\delta;t)$ is given as
\begin{equation}
 P(\delta;t)=\int \prod_{{\cal F}_i\neq\delta} d{\cal F}_i\ P(\calF;t),
\end{equation}
and similarly the one-point PDF of the dimensionless velocity 
divergence, $P(\theta;t)$, is
\begin{equation}
 P(\theta;t)=\int\prod_{{\cal F}_i\neq\theta} d{\cal F}_i\ P(\calF;t),
\end{equation}
where $\theta\equiv\nabla\cdot\bfv/(aH)$.

In general, a statistical characterization of the large-scale structure
is coordinate-dependent. 
Physically, there are at least two meaningful coordinates, i.e., the
Lagrangian and the Eulerian coordinates. 
While the Eulerian coordinate is fixed on a comoving frame, the
Lagrangian coordinate is fixed on fluid particles and follows the motion
of the fluid flow. 
Hence, as time goes on, high density regions in the Lagrangian space
occupy larger volume than those in the Eulerian space.  
We thus consider both the Lagrangian PDF $\probL$ and the Eulerian PDF
$\probE$, defined in the Lagrangian and the Eulerian coordinates,
$\bfq$ and  $\bfx$, respectively. 
The corresponding expectation values, $\langle\cdots\rangleL$ and
$\langle\cdots\rangleE$ are also introduced.

In the following subsection, we first consider the Lagrangian PDF and
derive the evolution equation. 
Then we derive the evolution equation for Eulerian PDF in section
\ref{subsec: derivation_Eul}. 
The evolution equations derived here are not yet closed because 
of the unknown functions. 
In section \ref{subsec: LocalApprox} we discuss an approximate treatment
using the local dynamics model, which enables us to obtain a closed form 
of evolution equation and to construct a consistent solution. 
%
%
%
%
%
%
%
%
%
\subsection{Lagrangian one-point PDF} 
\label{subsec:  derivation_Lag}
%
%
%
%
%
%
%
%
%
To derive the evolution equation for the Lagrangian one-point PDF, we
introduce an arbitrary function of local density, $g(\delta)$, and
consider the evolution of its expectation value, 
$\langle g(\delta(\bfq,t))\rangleL$ evaluated in a Lagrangian frame. 
The differentiation of this expectation value with respect to time
becomes 
\begin{eqnarray}
\frac{\partial}{\partial t}\langle g(\delta(\bfq,t))\rangleL
  =  \frac{\partial}{\partial t}\int\prod_i d{\cal F}_i\ 
 g(\delta)\,\,\probL(\calF;t) 
  =  \int d\delta\ g(\delta)\,\,\frac{\partial}{\partial t}\,
\probL(\delta;t), 
\label{dPLdt}
\end{eqnarray}
since the explicit time dependence of the quantity 
$\langle g(\delta)\rangleL$ only appears through the Lagrangian PDF. 
In the second equation, the integration is performed over the variable 
$\calF$ except for the local density $\delta$. 
On the other hand, the function $g$ implicitly depends on time through 
the evolution of Lagrangian local density $\delta(\bfq,t)$. 
Denoting the Lagrangian time derivative by 
$d/dt\equiv\partial/\partial t+\bfv/a\cdot\nabla$, 
the expectation value of the quantity $dg/dt$ becomes 
\begin{eqnarray}
\left\langle \frac{d}{d t}\,\,g(\delta(\bfq,t))\right\rangleL
=  \left\langle \frac{dg}{d \delta}\,\,\frac{d\delta}{dt}\right\rangleL
=  \int\prod_i d{\cal F}_i\ \frac{d g(\delta)}{d \delta}\,\,
        \frac{d\delta}{dt}\,\,\probL(\calF;t).   
\label{dgdt}
\end{eqnarray}
The right-hand-side of the above equation can be expressed by
integrating by part as 
\begin{eqnarray}
\int\prod_i d{\cal F}_i\ \,\,\frac{d g(\delta)}{d \delta}\,\,
\frac{d\delta}{dt}\,\,\probL(\calF;t) 
  &=& - \int\prod_i d{\cal F}_i\ g(\delta)\,\,
  \frac{\partial}{\partial \delta}\,
\left\{\,\,  \frac{d\delta}{dt}\,\,\probL(\calF;t) \,\,\right\}   
\nonumber \\
  &=& - \int d\delta\ \,\,g(\delta)\,\,\frac{\partial}{\partial \delta}\,
\left\{\,\, \left[\frac{d\delta}{dt}\right]_{\delta}\probL(\delta;t)
\,\,\right\}.   
\label{dPLddelta}
\end{eqnarray}
Here, the quantity $[A]_B$ denotes the conditional mean of $A$ for a
given value of $B$: 
\begin{equation}
  [A]_B \equiv \int \prod_{{\cal F}_i\neq B} d{\cal F}_i\ A\,\,P(\calF|B) 
  \label{def_of_[A]_B}
\end{equation}
with the function $P(\calF|B)$ being the conditional PDF for a given
$B$, i.e., $P(\calF|B)=P(\calF)/P(B)$.

Now, recalling the fact that $g(\delta)$ is an arbitrary function of
local density $\delta$, equation (\ref{dPLdt}) must be equivalent to
equation (\ref{dgdt}) in the Lagrange frame. 
The comparison between equation (\ref{dPLdt}) and equation
(\ref{dPLddelta}) then leads to the following evolution equation: 
\begin{equation}
\frac{\partial}{\partial t}\probL(\delta;t)+\frac{\partial}{\partial
 \delta}\left(\left[\frac{d\delta}{dt}\right]_\delta 
\probL(\delta;t)\right) = 0. 
\label{1ptLagdelta}
\end{equation}
Similarly, one can derive the evolution equation for the Lagrangian 
PDF of the velocity divergence $\probL(\theta;t)$: 
\begin{equation}
 \frac{\partial}{\partial t}\probL(\theta;t)+\frac{\partial}{\partial
 \theta}\left(\left[\frac{d\theta}{dt}\right]_\theta \probL(\theta;t)\right) 
= 0. 
\label{1ptLagtheta}
\end{equation}
%
%
%
%
%
%
%
%
%
\subsection{Eulerian one-point PDF}
\label{subsec: derivation_Eul}
%
%
%
%
%
%
%
%
%
The evolution equation for the Eulerian one-point PDFs can also be
derived by repeating the same procedure as above, but the resultant
expressions are slightly different from those of the Lagrangian PDFs. 
The time derivative of the expectation value
$\langle\,g(\delta)\,\rangleE$ becomes
\begin{equation}
\frac{\partial}{\partial t}\langle g(\delta(\bfx,t))\rangleE
=\int d\delta\ g(\delta)\frac{\partial}{\partial t}\probE(\delta;t).
\label{eq: partial_g_partial_t}
\end{equation}
As for the expectation value of $\partial g/\partial t$,    
with a help of the Lagrangian time derivative, we obtain
\begin{eqnarray}
\left\langle\frac{\partial}{\partial t}\,\,g(\delta(\bfx,t))\right\rangleE
 = \left\langle\frac{dg}{d\delta}\frac{\partial\delta}{\partial
 t}(\bfx,t)\right\rangleE 
=\left\langle\frac{dg}{d\delta}\frac{d\delta}{dt}(\bfx,t)\right\rangleE
-\left\langle\frac{dg}{d\delta}\frac{1}{a} \bfv\cdot \nabla\delta(\bfx,t)
\right\rangleE.
\label{delgdelt_E}
\end{eqnarray}
In the above expression, the first term in the second equation reduces
to the same expression as in equation (\ref{dPLddelta}) just replacing
the Lagrangian PDF with the Eulerian PDF. 
The second term in the second equation is further rewritten as 
\begin{eqnarray}
  \left\langle\frac{dg}{d\delta}\,\,
\frac{1}{a}\,\,\bfv\cdot \nabla\delta(\bfx,t)
\right\rangleE &=&
\left\langle \frac{1}{a}\,\,
\bfv\cdot\nabla g(\delta(\bfx,t))\right\rangleE
\nonumber\\
&=&\frac{1}{a}\,
\left\langle\nabla\cdot\{\bfv\,g(\delta(\bfx,t)\}\right\rangleE
\,\,-\,\,H
\left\langle \theta\,g(\delta(\bfx,t))\right\rangleE
\nonumber \\
&=& -H\,\int \prod_i\,d{\cal F}_i\,\,g(\delta)\,\theta\,\probE(\calF;t),
\end{eqnarray}
where we have used the fact that the first term in the second line
vanishes because of the isotropy. 
Then, using the definition of the conditional mean (\ref{def_of_[A]_B}),
equation (\ref{delgdelt_E}) can be summarized as follows: 
\begin{eqnarray}
\left\langle\frac{\partial}{\partial t}g(\delta(\bfx,t))\right\rangleE
 & = & \int d\delta\ g(\delta)\left\{ -\frac{\partial}{\partial\delta}\left(\left[\frac{d\delta}{dt}(t)
\right]_{\delta}\probE(\delta;t)\right) + H[\theta(t)]_{\delta}\probE(\delta;t)\right\}.
\end{eqnarray}
Hence, the relation 
$\langle \partial g/\partial t\rangleE=\partial\langle g(\delta)\rangleE/\partial t$, 
not the equation 
$\langle dg/dt\rangleE=\partial\langle g(\delta)\rangleE/\partial t$,
leads to the evolution equation for the Eulerian one-point PDF
$P(\delta,t)$. 
From equations (\ref{eq: partial_g_partial_t}) and (\ref{delgdelt_E}),
we obtain 
\begin{equation}
 \frac{\partial}{\partial t}\probE(\delta;t)+\frac{\partial}{\partial\delta}\left(\left[\frac{d\delta}
{dt}(t)\right]_{\delta}\probE(\delta;t)\right)=H[\theta(t)]_{\delta}\probE(\delta;t).
\label{1ptEuldelta}
\end{equation}
Comparing equation (\ref{1ptEuldelta}) with equation
(\ref{1ptLagdelta}), the only difference between the Eulerian and the 
Lagrangian PDF is the term $H[\theta]_\delta \probE$ in the
right-hand-side of the above equation, which represents the change of
the measure along the fluid trajectory in the Eulerian coordinate.

Finally, the evolution equation of the Eulerian one-point PDF
$\probE(\theta;t)$ is also derived in a similar way:
\begin{equation}
 \frac{\partial}{\partial t}\probE(\theta;t)+\frac{\partial}{\partial\theta}\left(\left[\frac{d\theta}
{dt}(t)\right]_{\theta}\probE(\theta;t)\right)=H\theta \probE(\theta;t).
\label{1ptEultheta}
\end{equation}
Note that the zero-mean of the velocity divergence $\langle\theta\rangleE=0$ 
is always guaranteed from equation (\ref{1ptEultheta}), which is easily shown 
by integrating the above equation over $\theta$ directly.
%
%
%
%
%
%
%
%
%
  \subsection{Local approximation}
\label{subsec: LocalApprox}
%
%
%
%
%
%
%
%
%
The evolution equations in the previous subsection are rather general
and in deriving these we did not specify the dynamics of fluid
evolution. 
In this sense, they are not closed until functional forms of the
conditional means $\left[d\delta/dt\right]_\delta$,
$\left[\theta\right]_\delta$ and $\left[d\theta/dt\right]_\theta$ are
specified. 
In other words, these equations require the additional equations
governing the evolution of the conditional means. 
However, an attempt to obtain the closed set of evolution equations
suffers from serious mathematical difficulty, so-called {\it closure
problem}, which is well-known in the subject of fluid mechanics and/or
plasma physics \citep[e.g.,][]{CCK1989,GK1993}. 
Similar to the BBGKY hierarchy, if one derives the evolution equations
for the conditional means, there appear new unknown conditional means. 
Thus, we must further repeat the derivation of evolution equation for
the new quantities. Continuing this operation, one could finally obtain
the infinite chain of the evolution equations, which is generally
intractable.

Instead of the exact analysis tackling the difficult problem, we rather
focus on an approximate treatment of the evolution of one-point PDFs,
where the solutions for the evolution equations are consistently
constructed. 
To implement this, we adopt the {\it local approximation} that the
evolution of the local density $\delta$ and the velocity-divergence
$\theta$ is described by the Lagrangian dynamics with finite degrees of 
freedom, whose initial conditions are characterized by the initial
parameters, $\bfp=(p_1,p_2,\cdots p_n)$, given at the same Lagrangian
coordinate. 
As will be explicitly shown in the next section, for instance, if the
spherical collapse model is adopted as Lagrangian local dynamics, the
evolution of local density is characterized by the single variable, 
which can be set as the linearly extrapolated density fluctuation,
$\delta_l$. 
If adopting the ellipsoidal collapse model, the dynamical degrees of
freedom reduce to three, representing the principal axes of initial
homogeneous ellipsoid, $\lambda_1$, $\lambda_2$ and $\lambda_3$. 
Thus, in this approximation, the density fluctuations can be expressed
as $\delta=f(\bfp,t)$, and using this expression, the
velocity-divergence is given by $\theta=-(df/dt)/H(1+f)$ from the
equation of continuity (\ref{continuity}). 
Within the local approximation, provided the initial distribution
function $\probI(\bfp)$, the form of the conditional means can be
completely specified and it can be expressed in terms of the functions
$\probI(\bfp)$ and $f(\bfp,t)$.

Let us first consider the Lagrangian PDF. 
In this case, the formal expressions for the conditional means 
$\left[d\delta/dt\right]_\delta$ and $\left[d\theta/dt\right]_\theta$
are given by 
\begin{eqnarray}
&& \left[\frac{d\delta}{dt}\right]_\delta=\frac{1}{\probL(\delta;t)}
\int \prod_i\,dp_i\,\,\probI(\bfp)\,\,\frac{d f(\bfp,t)}{d t}\,\,\delta_D(\delta-f(\bfp,t)),
\label{ddeltadt}
\\
&& \left[\frac{d\theta}{dt}\right]_\theta=\frac{1}{\probL(\theta;t)}
\int \prod_i dp_i\,\,\probI(\bfp)\,\,\frac{d}{dt}
\left\{-\frac{1}{H(1+f)}\frac{d f}{d t}\right\}
\,\,\delta_D\left(\theta+\frac{1}{H(1+f)}\frac{d f}{d t}\right).
\label{dthetadt}
\end{eqnarray}
With these expressions, the evolution equations (\ref{1ptLagdelta}) and 
(\ref{1ptLagtheta}) become a closed form and the 
consistent solutions can be constructed as follows: 
\begin{eqnarray}
& \probL(\delta;t)=&\int \prod_i dp_i\,\,\probI(\bfp)\,\,\delta_D(\delta-f(\bfp,t)),
\label{LagPDFdelta}
\\
& \probL(\theta;t)=&\int \prod_i dp_i\,\,\probI(\bfp)\,\,\delta_D\left(\theta+
\frac{1}{H\{1+f(\bfp,t)\}}\frac{d f(\bfp,t)}{d t}\right). 
\label{LagPDFtheta}
\end{eqnarray}
The proof that the above equations indeed satisfy the evolution equations 
(\ref{1ptLagdelta}) and (\ref{1ptLagtheta}) can be easily shown by
differentiating equations (\ref{LagPDFdelta}) and (\ref{LagPDFtheta})
with respect to time. 
For the PDF of the local density, one has 
\begin{eqnarray}
 \frac{\partial}{\partial t}\probL(\delta;t)&=& \int \prod_i dp_i\,\,
\probI(\bfp)\,\frac{\partial}{\partial t}\,\delta_D(\delta-f(\bfp,t))
\nonumber\\
&=& \int \prod_i dp_i\,\,
\probI(\bfp)\,\left(-\frac{d f}{dt}\right)\,
\frac{\partial}{\partial \delta}\,\delta_D(\delta-f(\bfp,t)) 
\nonumber 
\end{eqnarray}
from the property of the Dirac's delta function. 
In the above equation, the operator $\partial/\partial \delta$ in the
last line can be factored out and one can use the expression of
conditional mean (\ref{ddeltadt}). 
Then, the time derivative of the one-point PDF $\probL(\delta;t)$ is
rewritten as 
\begin{eqnarray}
  \frac{\partial}{\partial t}\probL(\delta;t) = - 
\frac{\partial}{\partial \delta}\,\left\{\,\left[\frac{d\delta}{dt}\right]_{\delta}\,
\probL(\delta;t)\,\right\}, 
\end{eqnarray}
which coincides with the evolution equation (\ref{1ptLagdelta}). 
As for the velocity-divergence PDF $\probL(\theta;t)$, we consistently
recover the evolution equation (\ref{1ptLagtheta}) with a help of
equation (\ref{dthetadt}): 
\begin{eqnarray}
 \frac{\partial}{\partial t}\probL(\theta;t)&=& \int \prod_i dp_i\,\,
\probI(\bfp)\,\,\frac{\partial}{\partial t}\,\,
\delta_D\left(\theta+\frac{1}{H(1+f)}\frac{d f}{d t}\right)
\nonumber\\
&=& \int \prod_i dp_i\,\,
\probI(\bfp)\,\,\frac{d}{d t}\,\left\{\frac{1}{H(1+f)}\frac{d f}{d t}\right\}
\,\,\frac{\partial}{\partial \theta}\delta_D\left(\theta+\frac{1}{H(1+f)}\frac{d f}{d t}\right)
\nonumber\\
&=&-\frac{\partial}{\partial \theta}
\left(\left[\frac{d\theta}{dt}\right]_\theta \probL(\theta;t)\right).
\end{eqnarray}

The approximate solution of the Eulerian one-point PDFs are also obtained 
similarly, 
but the factor $1/(1+\delta)$ must be convolved with the Lagrangian PDF 
due to the presence of inertial term 
(r.h.s of eqs.[\ref{1ptEuldelta}][\ref{1ptEultheta}]): 
\begin{eqnarray}
& \probE(\delta;t)=&\frac{1}{1+\delta}\int
 \prod_i dp_i\,\,\probI(\bfp)\,\,\delta_D(\delta-f(\bfp,t)), 
\\
& \probE(\theta;t)=&\int \prod_i dp_i\,\,
\frac{\probI(\bfp)}{1+f}\,\,\delta_D\left(\theta+\frac{1}{H\{1+f(\bfp,t)\}}
  \,\frac{d f}{d t}(\bfp,t)\right).
\end{eqnarray}
Note, however, that these PDFs do not satisfy the following conditions:
normalization condition $\langle 1\rangleE=1$ and zero means
$\langle\delta\rangleE=0$ and $\langle\theta\rangleE=0$. This fact 
simply reflects that the conservation of Eulerian volume cannot be always 
guaranteed, in contrast to the conservation of Lagrangian volume 
ensured by the mass conservation.  
As pointed out by \citet{FG1998a} \citep[see also][]{PS1997}, we 
re-scale the relation between $\delta$ and $f(\bfp,t)$ as follows: 
\begin{equation}
\delta=g(\bfp,t)\equiv N_E\,\{\,1+f(\bfp,t)\,\}\,-1~~;\,\,\,
N_E(t)\equiv\int \prod_i dp_i\,\,\frac{\probI(\bfp)}{1+f(\bfp,t)}.
\label{eq: function_g}
\end{equation}
Adopting this re-definition, the conditional means
$\left[d\delta/dt\right]_\delta$ and $\left[d\theta/dt\right]_\theta$
become 
\begin{eqnarray}
\left[\frac{d\delta}{dt}\right]_\delta &=&
\frac{1}{1+\delta}\,\,\frac{1}{\probE(\delta;t)}\int \prod_i dp_i\,\,\probI(\bfp)
\,\,\frac{dg}{dt}(\bfp,t)\,\,\delta_D(\delta-g(\bfp,t)),
\\
\left[\frac{d\theta}{dt}\right]_\theta &=&
\frac{1}{\probE(\theta;t)}\int \prod_i dp_i\,\,
\frac{\probI(\bfp)}{1+g}\,\,
\frac{dh}{dt}(\bfp,t)\,\,\delta_D\left(\theta-h(\bfp,t)\right),
\end{eqnarray}
where we define 
\begin{equation}
h(\bfp,t)\equiv -\frac{1}{H(1+g)}\frac{dg}{dt}.
\label{eq: function_h}
\end{equation}
Further, the conditional mean $[\theta]_{\delta}$ can be expressed as 
\begin{equation}
\left[\theta\right]_\delta = -\frac{1}{H(1+\delta)}\left[\frac{d\delta}{dt}\right]_\delta \label{dtheta/dt_with_delta}
\end{equation}
from the equation of continuity (\ref{continuity}).
Then, the solutions of equations (\ref{1ptEuldelta}) and
(\ref{1ptEultheta}) becomes 
\begin{eqnarray}
&\probE(\delta;t)=&\frac{1}{1+\delta}\int \prod_i dp_i\,\,\probI(\bfp)\,\,
\delta_D(\delta-g(\bfp,t)),
\label{EulPDFdelta}
\\
&\probE(\theta;t)=&\int \prod_i dp_i\,\,
\frac{\probI(\bfp)}{1+g}\,\,\delta_D\left(\theta-h(\bfp,t)
\right).
\label{EulPDFtheta}
\end{eqnarray}
One can easily show that equations (\ref{EulPDFdelta}) and
(\ref{EulPDFtheta}) satisfy the evolution equations (\ref{1ptEuldelta})
and (\ref{1ptEultheta}), with the correct normalization and the zero
mean.

The above solutions for Eulerian PDF can be regarded as a generalization
of the previous study based on the Zel'dovich approximation
\citep{Ko1994} and/or the spherical collapse model (\citealt{SG2001},
see also \citealt{PS1997}). 
Note, however, that the general expression of velocity-divergence PDF
$\probE(\theta;t)$ differs from the one obtained previously. 
While the factor $1/(1+\delta)$ is convolved in the integral in equation
(\ref{EulPDFtheta}), the resultant expression by \citet{Ko1994}
obviously omitted this factor. 
In our prescription, the PDF $\probE(\theta;t)$ is constructed from the
evolution equation, which basically relies on the equation of
continuity. 
This means that, even for the velocity-divergence PDF, the factor
$1/(1+\delta)$ is necessary to ensure the mass conservation. 
In fact, the expression (\ref{EulPDFtheta}) can also be obtained from
the joint PDF $\probE(\delta,\theta;t)$ integrating over the local
density $\delta$ (see eq.[\ref{PDFdeltatheta}]). 
Although the velocity-divergence PDF in the previous study has been
obtained in a rather phenomenological way, our present approach using
the evolution equations might be helpful in constructing the consistent
PDFs.

Nevertheless, even at this point, the solutions of evolution equations
should be regarded as formal expressions. 
In order to evaluate the PDFs explicitly, we need to specify the
Lagrangian local dynamics. 
In other words, the quantitative prediction for PDFs using the local
approximation crucially depends on the choice of the local dynamics. 
We will see in the section \ref{sec: demonstration} that the explicit
evaluation of $\probE(\delta;t)$ and $\probE(\theta;t)$ adopting the
spherical and the ellipsoidal collapse models shows several noticeable
differences. 
%
%
%
%
%
%
%
%
%
%
%
  \subsection{Joint PDF}
\label{subsec: Joint PDF}
%
%
%
%
%
%
%
%
%
%
%
So far, we have restricted our attention to the one-point PDF for the
single local variable, $\delta$ or $\theta$. 
In our treatment of the local approximation, the expressions for PDFs
are general forms irrespective of the Lagrangian local dynamics. 
However, it should be emphasized that if the local dynamics is
characterized by more than the two initial parameters, qualitative
behavior could be significantly changed from the local dynamics with
single degree of freedom. 
The point is that the relation between initial parameters and the
evolved result $\delta$ or $\theta$ cannot be described by a one-to-one
mapping. 
Accordingly, the relation between $\delta$ and $\theta$ becomes no
longer deterministic. 
Moreover, the failure of deterministic property also appears in the time
evolution of such local variables. 
It is therefore important to discuss the stochastic nature of $\delta$
and $\theta$ arising from the dynamical evolution. 
To characterize this, we consider the joint PDF.  
Within the local approximation, one can construct a consistent solution
of Eulerian joint PDF between $\delta$ and $\theta$ evaluated at the
same time, $\probE(\delta,\theta;t)$. 
Further, the Lagrangian joint PDF for the density field evaluated at the
same Lagrangian position but at the different times,
$\probL(\delta,t;\delta',t')$ can also be obtained.

The evolution equation of $\probE(\delta,\theta;t)$ can be derived
through the expectation value of an arbitrary function
$g(\delta,\theta)$. 
Repeating the same procedure as described in section 
\ref{subsec: derivation_Eul}, we obtain 
\begin{eqnarray}
& {\displaystyle \frac{\partial}{\partial t}\langle 
g(\delta(\bfx,t),\theta(\bfx,t))\rangleE }&
=\int d\delta d\theta\ g(\delta,\theta)\frac{\partial}{\partial t}\probE(\delta,\theta;t),
\nonumber \\
& {\displaystyle \left\langle \frac{\partial}{\partial t}g(\delta(\bfx,t),
\,\theta(\bfx,t))\right\rangleE } &= 
\left\langle \frac{\partial g}{\partial \delta}\frac{d\delta}{dt}
+\frac{\partial g}{\partial \theta}\frac{d\theta}{dt}+H\theta g(\delta,\theta)\right\rangleE 
\nonumber\\
&&= \int d\delta d\theta\ g(\delta,\theta)
\left\{-\frac{\partial}{\partial \delta}
\left(\left[\frac{d\delta}{dt}\right]_{\delta,\theta}\probE(\delta,\theta;t)\right)\right. 
\nonumber\\
&&~~~~~~~~~~~~~ \left.-\frac{\partial}{\partial \theta}
\left(\left[\frac{d\theta}{dt}\right]_{\delta,\theta}\probE(\delta,\theta;t)\right)+
H\theta \probE(\delta,\theta;t)\right\}.
\nonumber
\end{eqnarray}
Then, these two equations lead to the evolution equation for
$\probE(\delta,\theta;t)$: 
\begin{equation}
 \frac{\partial}{\partial t}\probE(\delta,\theta;t)-H\theta\frac{\partial}{\partial \delta}\left\{(1+
\delta)\probE(\delta,\theta;t)\right\}+\frac{\partial}{\partial \theta}\left(\left[\frac{d\theta}{dt}
\right]_{\delta,\theta}\probE(\delta,\theta;t)\right)=H\theta \probE(\delta,\theta;t), 
\label{jointdeltatheta}
\end{equation}
where we used the relation
$[d\delta/dt]_{\delta,\theta}=-H \theta(1+\delta)$.

The construction of the consistent solution in the local approximation
is almost parallel to the case of the Eulerian one-point PDF in section
\ref{subsec: LocalApprox}. 
The formal expression of the conditional mean
$[d\theta/dt]_{\delta,\theta}$ is 
\begin{equation}
 \left[\frac{d\theta}{dt}\right]_{\delta,\theta}=
 \frac{1}{\probE(\delta,\theta;t)}\frac{1}{1+\delta}\int \prod dp_i\ \,\,
 \probI(\bfp)\,\,\frac{d h}{d t}\,\,\delta_D(\delta-g(\bfp,t))\delta_D(\theta-h(\bfp,t)),
\end{equation}
and the corresponding solution of equation (\ref{jointdeltatheta})
becomes 
\begin{equation}
 \probE(\delta,\theta;t)=\frac{1}{1+\delta}\int \prod_i dp_i\,\, 
 \probI(\bfp)\,\,\delta_D(\delta-g(\bfp,t))\delta_D(\theta-h(\bfp,t)). 
\label{PDFdeltatheta}
\end{equation}
Note again that the above solution exactly recovers the one-point PDFs 
$\probE(\delta;t)$ and $\probE(\theta;t)$ integrating equation
(\ref{PDFdeltatheta}) over $\theta$ and $\delta$, respectively 
(see eqs.[\ref{EulPDFdelta}][\ref{EulPDFtheta}]).

The evolution equation for Lagrangian joint PDF is also obtained by the
time derivative of the expectation value of the arbitrary function,
$g(\delta(\bfq,t),\delta(\bfq,t'))$ as described in section 
\ref{subsec: derivation_Lag}: 
\begin{eqnarray}
&{\displaystyle \frac{d}{d t}
\left\langle g(\delta(\bfq,t),\delta(\bfq,t'))\right\rangle }&
= \int d\delta d\delta' g(\delta,\delta')
\frac{\partial}{\partial t}\probL(\delta,t;\delta',t'),
\nonumber \\
& {\displaystyle 
\left\langle\frac{d}{d t} g(\delta(\bfq,t),\delta(\bfq,t'))\right\rangle }&
= \left\langle \frac{\partial}{\partial\delta(\bfq,t)}g(\delta(\bfq,t),\delta(\bfq,t'))\frac{d
\delta(\bfq,t)}{dt}\right\rangle 
\nonumber \\
&&= -\int d\delta d\delta'g(\delta,\delta')\frac{\partial}{\partial\delta}\left(\left[\frac{d\delta}{dt}
\right]_{\delta,\delta'}\probL(\delta,t;\delta',t')\right).
\nonumber
\end{eqnarray}
The evolution equation of Lagrangian joint PDF is 
\begin{equation}
 \frac{\partial}{\partial t}\probL(\delta,t;\delta',t')+
\frac{\partial}{\partial\delta}\left(\left[\frac{d\delta}{dt}\right]_{\delta,\delta'}
\probL(\delta,t;\delta',t')\right)=0.
\label{eq: evolve_jointPDF_Lag}
\end{equation}
The conditional mean in the local approximation is expressed as 
\begin{eqnarray}
\left[\frac{d\delta}{dt}\right]_{\delta,\delta'}&=&
\frac{1}{\probL(\delta,t;\delta',t')}\int \prod_i dp_i\,\,\probI(\bfp)\,\,
\frac{d f(\bfp,t)}{d t}\,\,\delta_D(\delta-f(\bfp,t))\,\delta_D(\delta'-f(\bfp,t')).
\nonumber
\end{eqnarray}
Recalling that the joint PDF satisfying the evolution equation 
(\ref{eq: evolve_jointPDF_Lag}) should be invariant under the
transformation, $(\delta,t)\leftrightarrow (\delta',t')$, the solution
consistent with the boundary condition 
$\probL(\delta,t';\delta',t')=\probL(\delta;t')\delta_D(\delta-\delta')$
becomes 
\begin{eqnarray}
\probL(\delta,t;\delta',t')&=& 
\int \prod_i dp_i\,\,\probI(\bfp)\,\,\delta_D(\delta-f(\bfp,t))\,
\delta_D(\delta'-f(\bfp,t')). 
\label{jointpdf}
\end{eqnarray}
Notice that if the local Lagrangian dynamics is described by a single
parameter, the integral over the initial parameter $p_1$ in equation
(\ref{jointpdf}) can be formally performed. 
The resultant expression includes Dirac's delta function, leading to the
one-to-one mapping between $\delta$ and $\delta'$. 
On the other hand, in cases with the multivariate initial parameters,
one cannot generally perform the above integral and the Dirac's delta
function is not factored out, leading to the stochastic nature of local
density fields.

One might further consider the evolution of Eulerian joint PDF 
$\probE(\delta,t;\delta',t')$, which has been indeed shown in N-body
simulations by \citet{Ka2001}. 
The derivation of evolution equation itself is an easy task, but, the
formal solution in the local approximation suffers from difficulties due
to the presence of advective term, which might be related to an
important effect on the non-local nature of fluid flows. 
Since even the Lagrangian joint PDF shows several important features,
one can expect that the qualitatively similar behavior to the N-body
results can be seen from the Lagrangian joint PDF. 
Hence, we will postpone to analyze the Eulerian joint PDF 
$\probE(\delta,t;\delta',t')$ and instead focus on the Lagrangian joint
PDF $\probL(\delta,t';\delta',t')$. 
%
%
%
%
%
%
%
%
%
%
\section{DEMONSTRATION AND RESULTS} 
\label{sec: demonstration}
%
%
%
%
%
%
%
%
%
Now we are in a position to give an explicit evaluation of the one-point
PDF based on the local approximation discussed in the previous section. 
For an illustrative purpose, we adopt the spherical and the ellipsoidal
collapse models as simple and intuitive Lagrangian local dynamics. 
After briefly describing the basic equations of these models in section
\ref{subsec: SCA_ECA}, we compute the Eulerian one-point PDFs
$\probE(\delta)$ and $\probE(\theta)$ and discuss the qualitative
differences arising from the choice of Lagrangian local dynamics in
section \ref{subsec: results}. 
In particular the stochasticity in the multi-variate function of local
density or velocity-divergence is examined in detail by evaluating the
joint PDFs, $\probL(\delta,t;\delta',t')$ and $\probE(\delta;\theta)$
from the ellipsoidal collapse model. 
%
%
%
%
%
%
%
%
\subsection{Models for Lagrangian local dynamics} 
\label{subsec: SCA_ECA}
%
%
%
%
%
%
%
%
First consider the simplest case of the Lagrangian local dynamics, in
which the evolution of local quantities is determined by the mass inside
a sphere of radius $R$ collapsing via self-gravity: 
\begin{equation}
 \frac{d^2R}{dt^2}=-\frac{GM}{R^2} ~~~~~;~~
 M=\frac{4\pi}{3}\,\,\overline\rho\,\, R^3=\mbox{const},
\end{equation}
where $M$ is the mass inside the radius and $\overline{\rho}$ represents
the local density. 
This spherical collapse model can be re-expressed as the evolution
equation for density fluctuations $\delta$, given by 
$\delta=\overline{\rho}/\rho_{\rm m}-1=(a/R)^3-1$ as follows: 
\begin{equation}
 \frac{d^2\delta}{dt^2}+2H\frac{d\delta}{dt}-
 \frac{4}{3}\frac{1}{1+\delta}\left(\frac{d\delta}{dt}
\right)^2=\frac{3}{2}H^2\Omega_{\rm m}(1+\delta)\delta,\label{SCequation}
\end{equation}
with the quantity $\Omega_{\rm m}$ being the density parameter, 
$\Omega_{\rm m}\equiv8\pi G\,\rho_{\rm m}/(3H^2)$. 
As \citet{FG1998a} stated, this equation is regarded as a shear-less and
irrotational approximation of fluid equations, since one can derive the 
following equation from equations (\ref{continuity}) to (\ref{Poisson}) : 
\begin{equation}
 \frac{d^2\delta}{dt^2}+2H\frac{d\delta}{dt}-\frac{4}{3}\frac{1}{1+\delta}\left(\frac{d\delta}{dt}
\right)^2=H^2(1+\delta)\left(\frac{3}{2}\Omega_{\rm m}\delta+\sigma^{ij}\sigma_{ij}-\omega^{ij}\omega_{ij}
\right),
\end{equation}
with a help of the Lagrangian time derivative. 
Here the quantities $\sigma_{ij}$ and $\omega_{ij}$ respectively denote 
the shear and the rotation given by 
\begin{eqnarray}
\sigma_{ij}&=&\frac{1}{2aH}\left(\frac{\partial v_i}{\partial x_j}+\frac{\partial v_j}{\partial x_i}
\right)-\frac{1}{3}\,\,\theta\,\,\delta_{ij},\\
\omega_{ij}&=&\frac{1}{2aH}\left(\frac{\partial v_i}{\partial x_j}-\frac{\partial v_j}{\partial x_i}
\right). 
\end{eqnarray}
In the spherical collapse approximation (\ref{SCequation}), the density
fluctuations $\delta$ depend only on a single initial parameter
$\delta_l$, i.e., the linear fluctuation at an initial time, if the
decaying mode of the linear perturbation is neglected. 
Note that the spatial distribution of the initial density field is
randomly given and thereby the parameter $\delta_l$ is regarded as a
random variable. 
We assume that the initial parameter $\delta_l$ obeys a Gaussian
distribution: 
\begin{equation}
  \label{eq: gauss_PDF}
  \probI(\delta_l) = \frac{1}{\sqrt{2\pi}\,\,\sigma_l}\,\,
  e^{-(\delta_l/\sigma_l)^2/2}, 
\end{equation}
where the variable $\sigma_l$ means the rms fluctuation of the linear
density field $\delta_l$, i.e., $\sigma_l^2=\langle\delta_l^2\rangle$.

The ellipsoidal collapse approximation which is another Lagrangian local
dynamics, describes the evolution of the uniform density ellipsoid. 
In contrast to the spherical collapse model, the evolution of $\delta$
is governed by the dynamics of the half length of principal axes
$\alpha_i$ $(i=1,2,3)$ characterizing the density ellipsoid. 
According to \citet{BM1996}, we have 
\begin{equation}
 \frac{d^2}{dt^2}\alpha_i= -4\pi G\,\rho_m\,\alpha_i\left(\frac{1+\delta}
{3}+\frac{b_i}{2}\delta +\lambda_{{\rm ext},i}\right) \label{EllipAxis},
\end{equation}
\begin{equation}
 b_i=\alpha_1\alpha_2\alpha_3\int_0^\infty\frac{d\tau}{(\alpha_i^2+\tau)\prod_j(\alpha_j^2+\tau)^
{1/2}}-\frac{2}{3}, 
\end{equation}
and the relation between $\delta$ and $\alpha_i$ becomes 
\begin{equation}
 \delta=\frac{a^3}{\alpha_1\alpha_2\alpha_3}-1.
\label{eq: delta_ellipsoid}
\end{equation}
Here, the variable $\lambda_{{\rm ext},i}$ quantifies the external tidal
effect, which is required for the consistency with the Zel'dovich
approximation in a linear regime \citep{BM1996}: 
\begin{equation}
 \lambda_{{\rm ext},i}=\cases{\displaystyle 
D(t)\,\left(\lambda_i-\frac{\lambda_1+\lambda_2+\lambda_3}{3}\right) &; 
 linear external tide , \cr
                        \frac{5}{4}\,b_i &; nonlinear external tide ,}
\label{eq: external_tide}
\end{equation}
where $D(t)$ is the linear growth rate. 
The variables $\lambda_i$ represent the initial parameters of principal
axes, and in terms of these, the initial conditions reduce to 
\begin{eqnarray}
 \alpha_i(t_0)&=&a(t_0)\{1-D(t_0)\lambda_i\},
\label{eq: init_con(1)}
\\
 \frac{d}{dt}\alpha_i(t_0)&=&
\dot{a}(t_0)\{1-D(t_0)\lambda_i\}-a(t_0)\dot{D}(t_0)\lambda_i.
\label{eq: init_con(2)}
\end{eqnarray}
In contrast to the spherical collapse model, one can regard this model
as an approximation of the fluid equations taking account of the
influence of shear but neglecting the rotation: 
\begin{eqnarray}
&& \frac{d^2\delta}{dt^2}+2H\frac{d\delta}{dt}-\frac{4}{3}\frac{1}{1+\delta}\left(\frac{d\delta}{dt}
\right)^2=H^2(1+\delta)\left(\frac{3}{2}\Omega_{\rm m}\delta+\sigma^{ij}\sigma_{ij}\right); 
\label{eq: eom_delta_ellipsoid}
\\
 &&~~~~~~~~~~~~~~~
\sigma_{ij}=\frac{1}{3H}\left(3\frac{\dot{\alpha_i}}{\alpha_i}-\frac{\dot{\alpha_1}}{\alpha_1}-\frac
{\dot{\alpha_2}}{\alpha_2}-\frac{\dot{\alpha_3}}{\alpha_3}\right)\,\delta_{ij}.
\nonumber
\end{eqnarray}
Note that similar to the spherical collapse approximation, the three
initial parameters $\lambda_i$ are regarded as the random variables. 
When the initial condition of density field is assumed to be a Gaussian
random distribution, the expression for the initial parameter
distribution $\probI(\lambda_i)$ is analytically obtained as follows
\citep[e.g.,][]{D1970,BBKS1986}: 
\begin{equation}
 \probI(\lambda_i)=\frac{3375}{8\sqrt{5}\pi\sigma_l^6}\exp\left(-3\frac{I_1^2}{\sigma_l^2}+15\frac{I_2}
{2\sigma_l^2}\right)(\lambda_1-\lambda_2)(\lambda_2-\lambda_3)(\lambda_1-\lambda_3),
\label{DoroPDF}
\end{equation}
where the quantities $I_1$ and $I_2$ denote 
$I_1\equiv\lambda_1+\lambda_2+\lambda_3$ and 
$I_2\equiv\lambda_1\lambda_2+\lambda_2\lambda_3+\lambda_3\lambda_1$, 
respectively.

\begin{figure}
\epsscale{0.7}
 \plotone{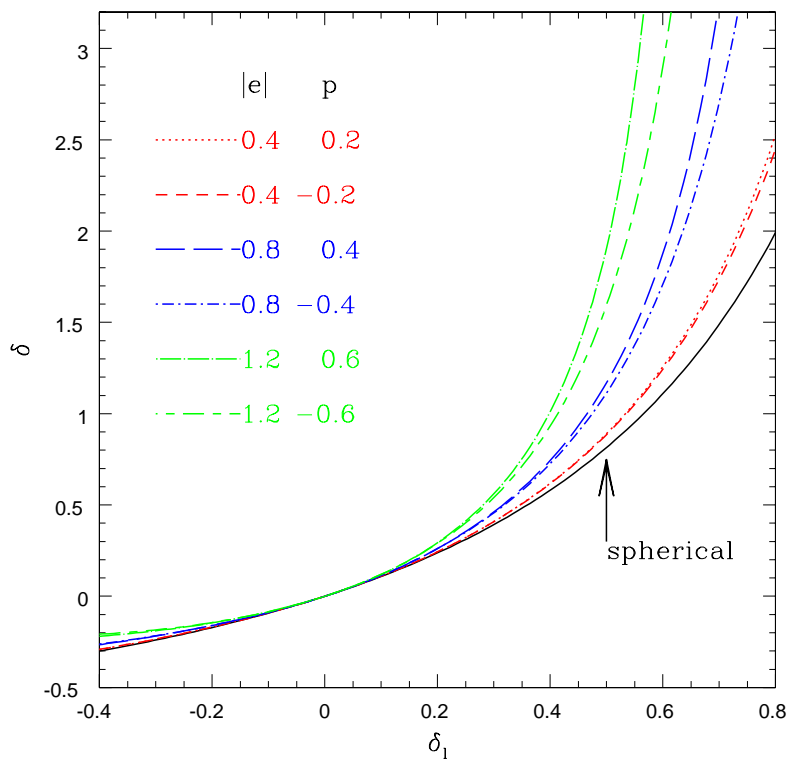}
 \figcaption{Evolution of local density from the ellipsoidal collapse model 
for various initial parameters, $(e,p)$, where $e$ is ellipticity and $p$ 
is prolateness, given by $e=(\lambda_1-\lambda_3)/(2\delta_l)$ and 
$p=(\lambda_1+\lambda_3-2\lambda_2)/(2\delta_l)$, respectively. 
The evolved results are then plotted as a function of linearly extrapolated 
density, $\delta_l=\lambda_1+\lambda_2+\lambda_3$. For comparison, we also 
plot the one-to-one local mapping from the spherical collapse model 
({\it solid}).
 \label{fig:evolve}}
\end{figure}

Based on these Lagrangian local models, we numerically calculate the
PDFs assuming the Einstein-de Sitter universe 
($\Omega_{\rm m}=1$, $\Omega_{\Lambda}=0$), in which the linear growth
rate $D$ is simply proportional to the scale factor $a$. 
For a better understanding of the later analysis, in 
Figure \ref{fig:evolve}, we plot the evolution of local density $\delta$ 
from the ellipsoidal collapse model for some initial conditions $(e,p)$ 
given by $e=(\lambda_1-\lambda_3)/(2\delta_l)$ and 
$p=(\lambda_1+\lambda_3-2\lambda_2)/(2\delta_l)$. The results are then 
depicted as a function of linearly extrapolated density 
$\delta_l=\lambda_1+\lambda_2+\lambda_3$ and are compared with 
the one from the one-to-one mapping of spherical collapse model ({\it solid}). 
Figure \ref{fig:evolve} shows that 
the local density of the ellipsoidal collapse model generally takes 
a larger value than that of the spherical collapse model. Further, 
the variety of evolved density for fixed $\delta_l$ 
suggests that a large amount of scatter will appear in the joint PDF 
$\probL(\delta,t;\delta',t')$ and the resultant one-point PDFs 
$\probE(\delta)$ and $\probE(\theta)$ cannot, in general, coincide 
with those obtained from the spherical collapse model. 
%
%
%
%
%
%
%
%
\subsection{Results}
\label{subsec: results}
%
%
%
%
%
%
%
In computing the PDFs from the above local collapse models, one may
practically encounter the case when the local density infinitely
diverges at finite elapsed time for some regions in the initial
parameter space, which has not been treated in previous section. 
To avoid the unphysical divergences, we must restrict the integral in
the PDFs to the initial parameter space $V(t)$, in which the local
density $\delta$ does not diverge. 
Indeed, this modification slightly affects the normalization condition
for both the Lagrangian and the Eulerian PDFs, i.e.,
$\langle\,1\,\rangle_{\scriptscriptstyle L,E}\neq1$. 
Although this does not alter any qualitative features of PDFs, we
consider some modifications to keep the correct normalization and
adopting this procedure in appendix \ref{appen: calculation_of_PDFs},
and the results for one-point and joint PDFs are presented below. 
Note, however, that the perturbation calculation discussed in
\ref{subsubsec: one-point_PDFs} is free from the serious divergences and
within the perturbative treatment, one can rigorously develop the local
approximation for PDFs. 
%
%
%
%
%
%
\subsubsection{One-point PDFs} 
\label{subsubsec: one-point_PDFs}
\begin{figure}

\vspace*{-1.5cm}

\epsscale{0.7}
 \plotone{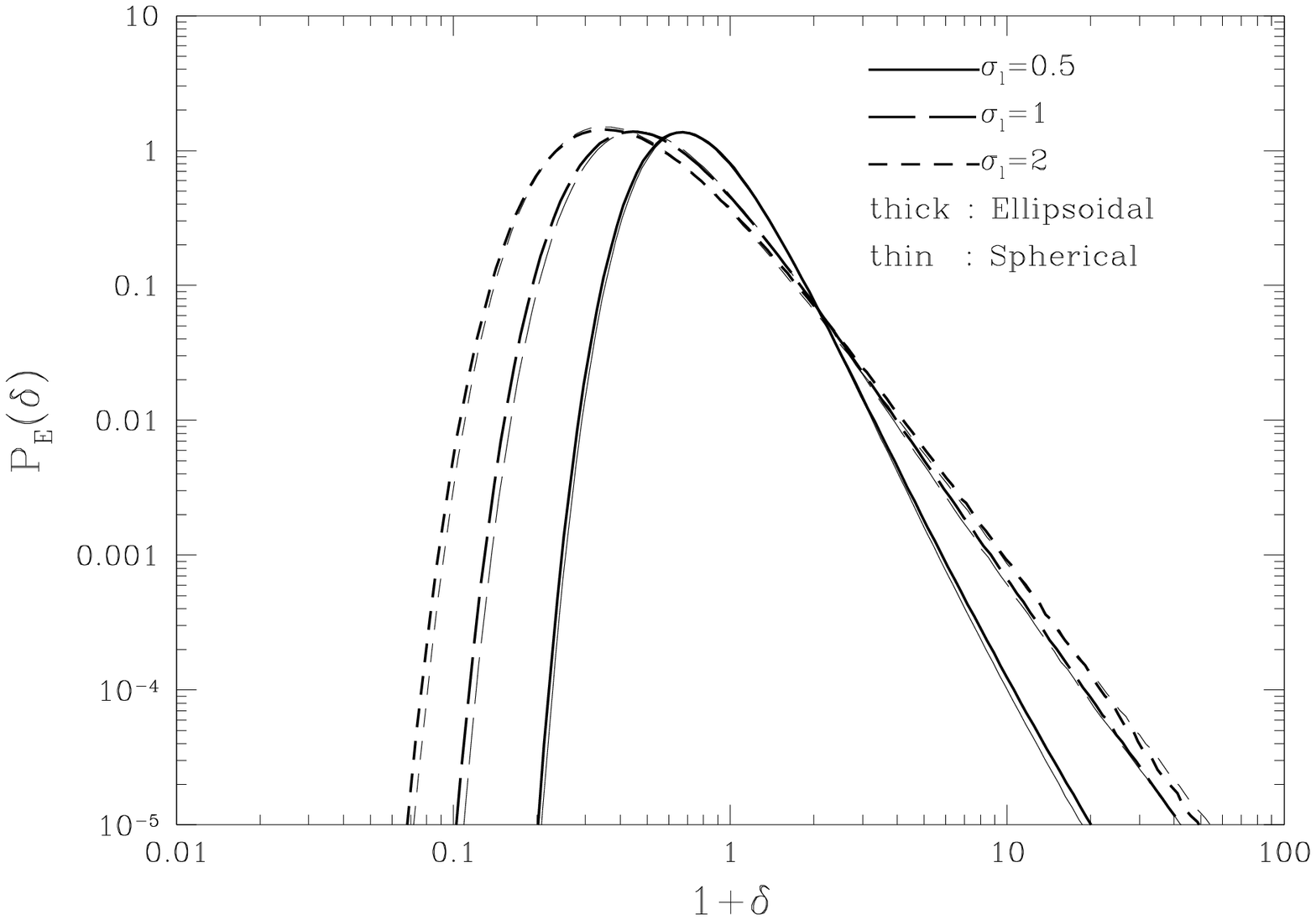}

\vspace*{-3.5cm}

 \plotone{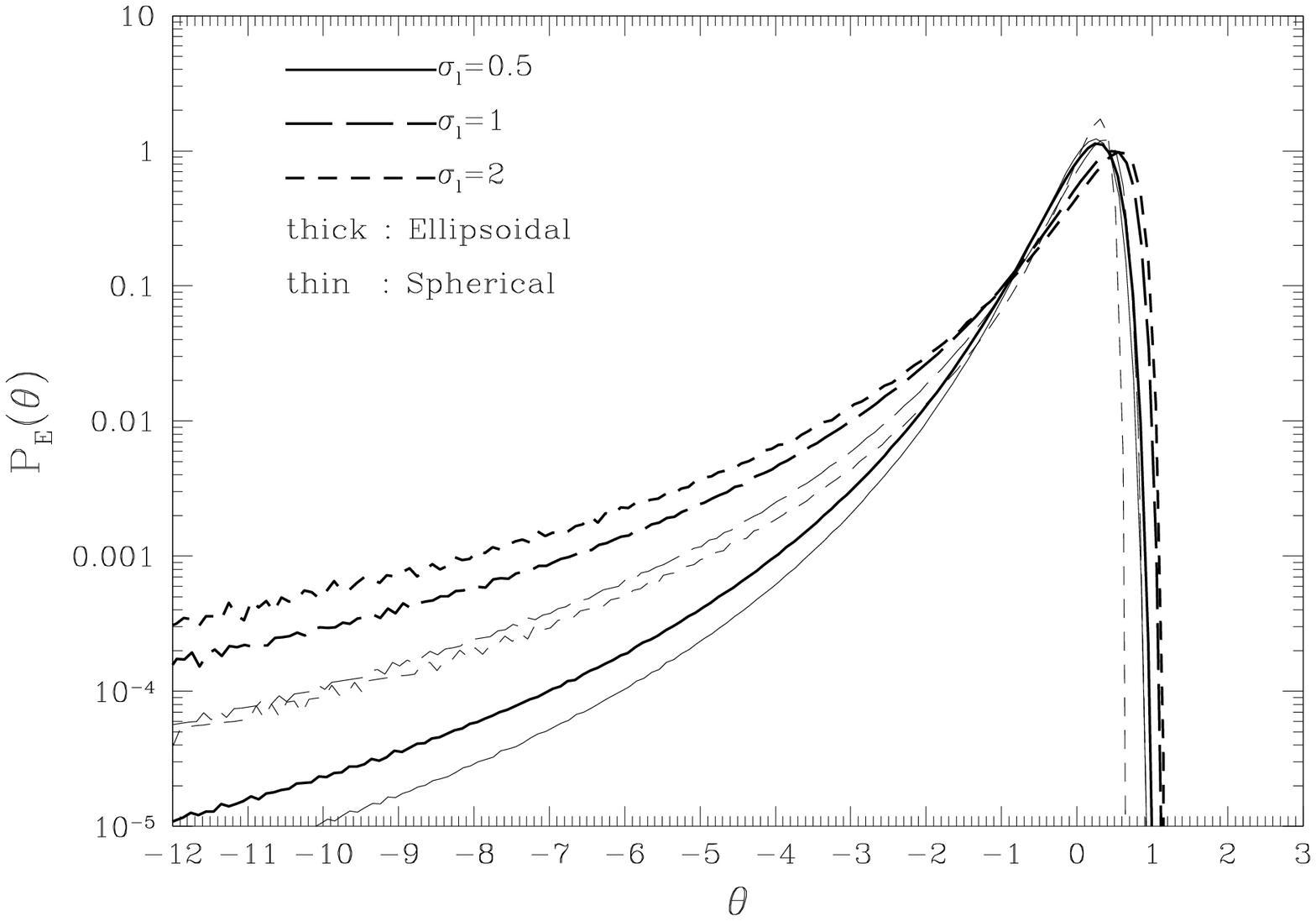}

\vspace*{-1.0cm}

 \figcaption{Eulerian one-point PDF, $\probE(\delta)$({\it top}) and 
  $\probE(\theta)$({\it bottom}) for the ellipsoidal collapse model with 
  linear external tide (thick lines) and for the spherical collapse model 
  (thin lines). 
  In each panel, solid, long-dashed and short-dashed lines represent the
  results in cases with the linear rms fluctuation $\sigma_l=0.5$, $1$
  and $2$, respectively. 
 \label{fig:1ptEulPDF}}
\end{figure}

%
%
%
%
%
Let us show the results of the one-point PDF. 
Figure \ref{fig:1ptEulPDF} plots the one-point Eulerian PDFs of the
local density ({\it top}) and the velocity-divergence ({\it bottom})
evaluated at various linear fluctuation values $\sigma_l$. 
In both panels, the thick lines represent the results obtained from the
ellipsoidal collapse model with linear external tide, while the thin
lines denote the PDFs from the spherical collapse models. 
In computing the PDFs, the Lagrangian local dynamics are numerically
solved with the various initial conditions $\bfp$ in the initial
parameter space $V$. 
Then, weighting by the PDF of the initial parameter $\probI(\bfp)$, the
PDFs are constructed by binning the evolved results of the density
$\delta$ and the velocity-divergence $\theta$, together with appropriate
convolution factors (see Appendix \ref{appen: calculation_of_PDFs}).

As expected, the overall behaviors of both PDFs in Figure
\ref{fig:1ptEulPDF} are qualitatively similar, irrespective of the
Lagrangian local models. 
As increasing the linear fluctuation value $\sigma_l$, while the density
PDF $\probE(\delta)$ extends over the high-density region $\delta\gg1$,
the velocity-divergence PDF $\probE(\theta)$ is negatively skewed and it
extends over the negative region $\theta \ll -1$. 
In looking at the differences in each local model, we readily observe
several remarkable features. 
First, the density PDFs computed from both the spherical and the 
ellipsoidal collapse models almost agree with each other. 
At first glance, this seems to contradict with a naive 
expectation from the local dynamics in Figure \ref{fig:evolve}.     
However, one might rather suspect that the agreement in density PDFs 
is accidental, due to the distribution of 
initial parameters $\probI(\lambda_i)$, 
which is, at least, consistent with a naive 
picture that joint PDF $\probL(\delta,t;\delta',t')$ from the 
ellipsoidal collapse model exhibits a large mount scatter 
and the mean relation between $\delta$ and $\delta'$ significantly 
deviates from one-to-one mapping of spherical collapse model 
(see Sec.\ref{subsubsec: joint_PDFs} and Fig.\ref{fig:jointPDF}). 
On the other hand, the velocity-divergence PDFs from the ellipsoidal 
collapse model exhibit 
longer non-Gaussian tails, compared with those obtained from the
spherical collapse model. 
The deviation between both models in PDF $\probE(\theta)$ becomes
significant as increasing the value $\sigma_l$. 
Interestingly, in the non-linear regime $\sigma_l\geq1$, tails of PDF
$\probE(\theta)$ from the spherical collapse model show the opposite
tendency, i.e., the amplitude decreases as increasing $\sigma_l$.

In order to characterize the qualitative behaviors more explicitly, we
perturbatively solve the evolution equations for both the spherical and
the ellipsoidal collapse models. 
The differences are then quantified evaluating the higher order moments
of one-point statistics for the local density and velocity-divergence. 
In appendix \ref{appen: perturbation}, based on the formal solution of
PDFs in section \ref{subsec: LocalApprox}, perturbative calculations of
local collapse models are briefly summarized and the solutions up to the
fifth order are presented. 
The resultant expressions for the higher order moments of density and
velocity-divergence are obtained as a series expansion of linear
variance $\sigma_l^2$, up to the two-loop order for the variance and the
one-loop order for the skewness and the kurtosis: 
\begin{eqnarray}
\sigma^2&\equiv&\langle\delta^2\rangleE=
\sigma_l^2 + s_{2,4} \,\,\sigma_l^4 + s_{2,6} \,\,\sigma_l^6+\cdots,
\label{eq: perturb_sigma}
\\
S_3&\equiv& \frac{\langle\delta^3\rangleE}{\sigma^4}
= S_{3,0} + S_{3,2} \,\,\sigma_l^2+\cdots,
\label{eq: perturb_S3}
\\
S_4&\equiv& \frac{\langle\delta^4\rangleE-3\sigma^4}{\sigma^6}
= S_{4,0} + S_{4,2} \,\,\sigma_l^2+\cdots
\label{eq: perturb_S4}
\end{eqnarray}
for the local density and 
\begin{eqnarray}
\sigma_{\theta}^2 &\equiv &\langle\theta^2\rangleE =
\sigma_l^2 + s_{2,4}^{\theta}\,\, \sigma_l^4 + s_{2,6}^{\theta}\,\, 
\sigma_l^6+\cdots, 
\label{eq: perturb_sigma_th}
\\
T_3 &\equiv &\frac{\langle\theta^3\rangleE}{\sigma_{\theta}^4}
= T_{3,0} + T_{3,2} \,\,\sigma_l^2+\cdots,
\label{eq: perturb_T3}
\\
T_4 &\equiv &\frac{\langle\theta^4\rangleE-3\sigma_{\theta}^4}{\sigma_{\theta}^6}
= T_{4,0} + T_{4,2} \,\,\sigma_l^2+\cdots
\label{eq: perturb_T4}
\end{eqnarray}
for the velocity-divergence. 
Then, all the coefficients in the above expansions yield the rigorous
fractional number and Table \ref{tbl : loop_correction} displays a
summary of the results. 
The calculation in spherical collapse model essentially reproduces the
non-smoothing results obtained by \citet{FG1998a,FG1998b}. 
Note, however, that the discrepancy has appeared in the higher order
correction of velocity-divergence 
\citep[c.f., eq.{[12]} with $\gamma=0$ of][]{FG1998b}. 
Perhaps, in computing the velocity-divergence moments, \citet{FG1998b}
incorrectly used the cumulant expansion formula for $\delta$ listed in
\citet{FG1998a}. 
Further, we suspect that they erroneously replaced the convolution
factor $1/(1+\delta)$ in Eulerian expectation value with
$1/(1+\theta)$. 
On the other hand, in our calculation, moments 
$\langle\theta^n\rangleE$ are rigorously computed according to the
definition (\ref{eq: moment_theta}), with a help of the
velocity-divergence PDF (\ref{EulPDFtheta}) with equation 
(\ref{eq: function_h}). 
Hence, the present calculation is at least consistent with the local
approximation in section \ref{subsec: LocalApprox} and we believe that
no serious error has appeared in present result.

\begin{figure}
\epsscale{1.0}
 \plottwo{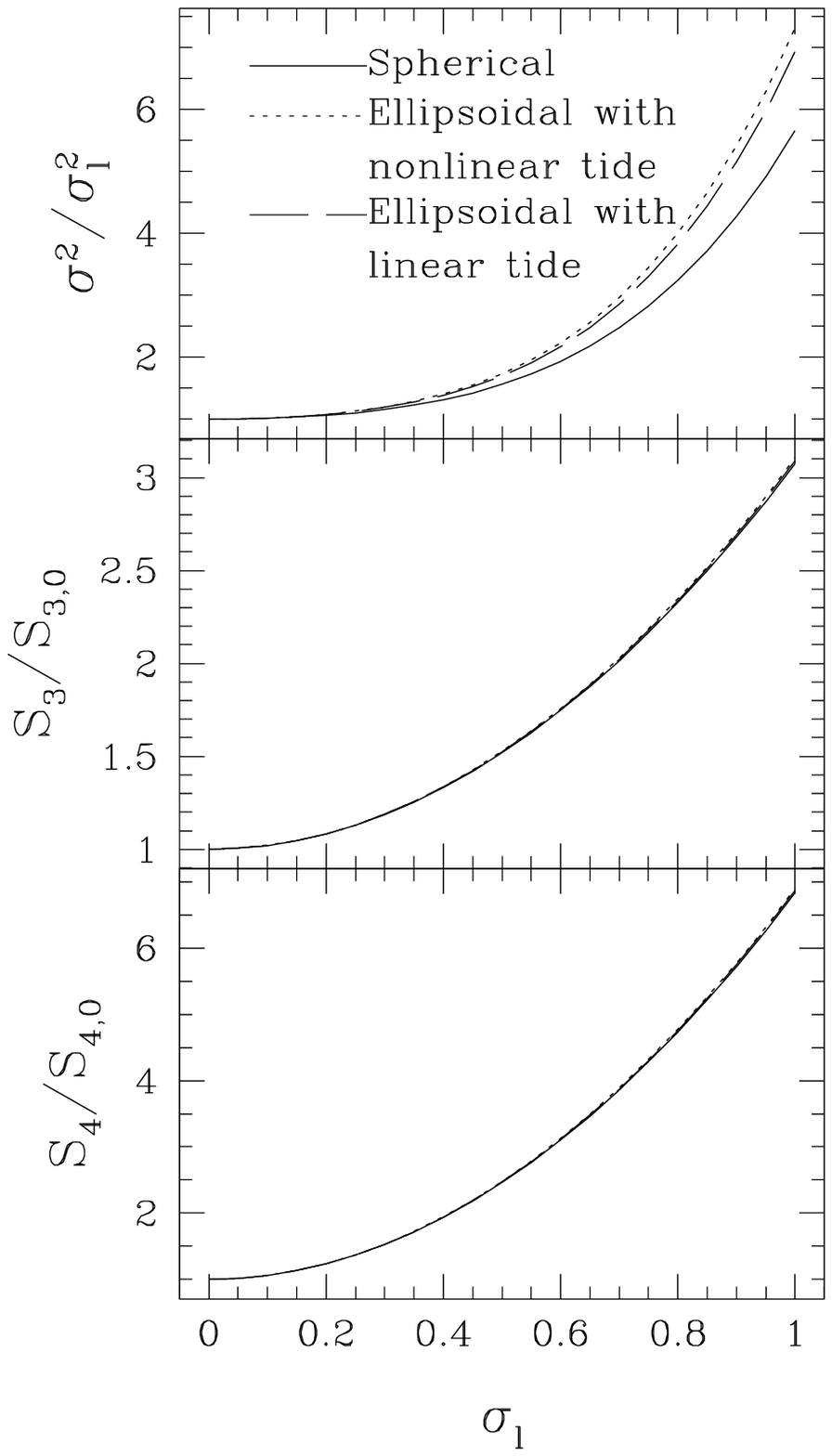}{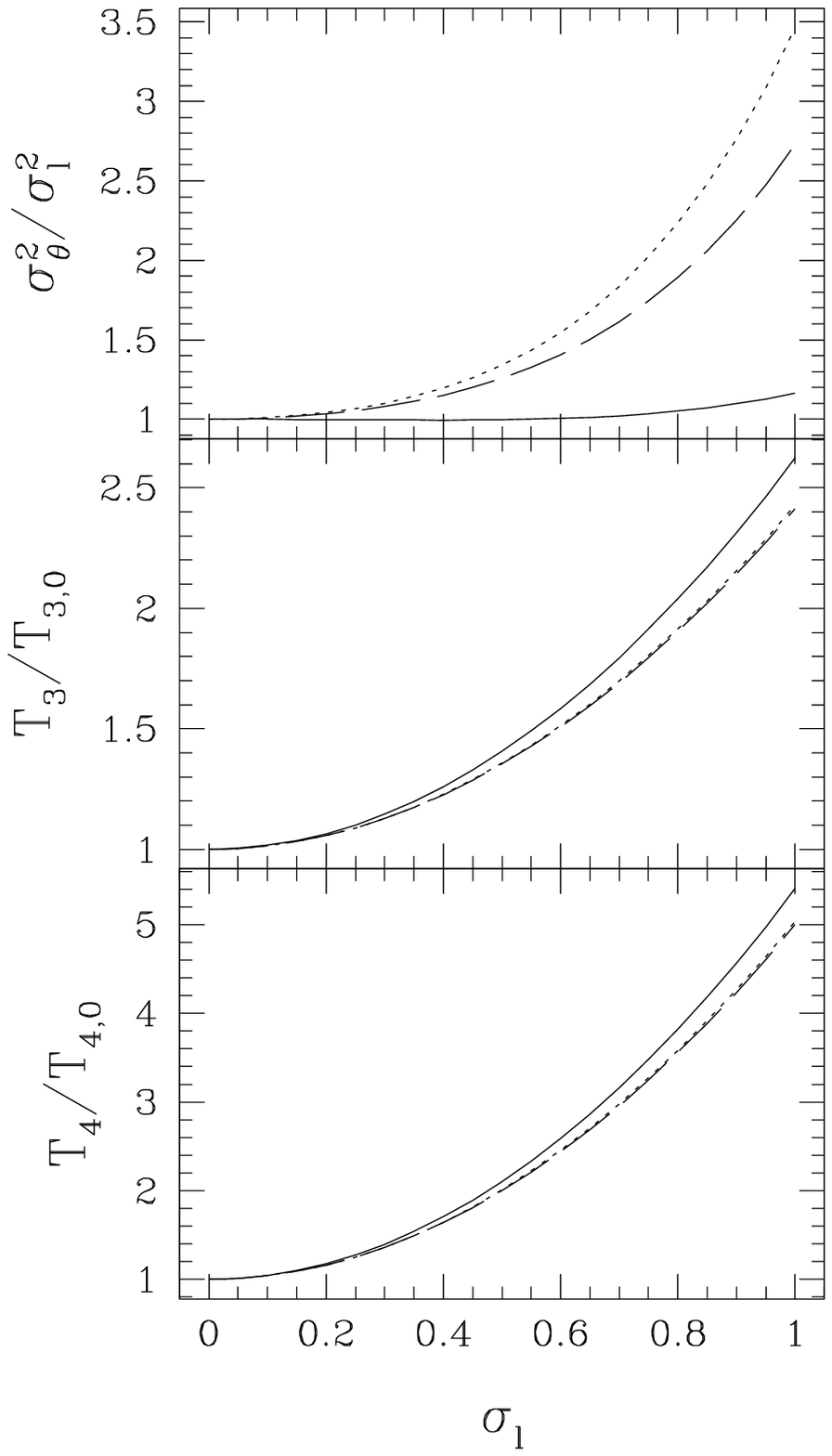}
 \figcaption{Departures of the variance, the skewness and the kurtosis
  of local density({\it left}) and the velocity-divergence({\it right})
  from the tree-level perturbation results. 
  While the variances $\sigma^2$ and $\sigma_{\theta}^2$ are calculated
  up to the sixth order of the linear rms fluctuation $\sigma_l$, the
  perturbation results for the skewness $S_3,\,T_3$  and the kurtosis
  $S_4,\,T_4$ are obtained up to the one-loop order, ${\cal O}(\sigma_l^2)$  
  (see Appendix \ref{appen: perturbation} in detail).   
  In each panel,  the dashed and the dotted lines indicate the results
  obtained from the ellipsoidal collapse model with linear external tide
  and nonlinear external tide, respectively. 
  For comparison, the results from the spherical collapse model are also
  shown in solid lines. 
 \label{fig:MomentDelta} }
\end{figure}

Apart from this discrepancy, one finds that the leading-order
(tree-level) calculation of skewness $S_{3,0}$ and kurtosis $S_{4,0}$ in
both models exactly coincides with each other. 
While the differences in the higher order correction for local density
are basically small, consistent with Figure \ref{fig:1ptEulPDF}, the
results in velocity-divergence exhibit a large difference, especially in
the variance $\sigma_{\theta}^2$. 
Figure \ref{fig:MomentDelta} summarizes the departure from the
leading-order perturbations for the variance({\it top}), the
skewness({\it middle}) and the kurtosis({\it bottom}), each of which is
normalized by the leading term. 
Clearly, the higher order corrections for variance $\sigma_{\theta}^2$
show the significant difference between the spherical and the
ellipsoidal collapse models, although the model dependence of the
external tide in ellipsoidal collapse is rather small. 
Remarkably, the one-loop correction $s_{2,4}^{\theta}$ is negative in
the spherical collapse model and thereby the quantity $\sigma_{\theta}$
does not monotonically increase. 
This behavior indeed matches with the non-monotonic behavior of
velocity-divergence PDF seen in Figure \ref{fig:1ptEulPDF}. 
In this sense, the perturbation results successfully explain the
numerical results of PDF. 
This fact further indicates that in a Gaussian initial condition,  
the influence of non-sphericity or
effect of shear could be negligible in the one-point statistics of local
density, while it alters the shape of the PDF
$\probE(\theta)$, which might be a natural outcome of the multivariate
local approximation. 
%
%
%
%
%
%
%
%
\subsubsection{Joint PDFs} 
\label{subsubsec: joint_PDFs}
\begin{figure}
\plotone{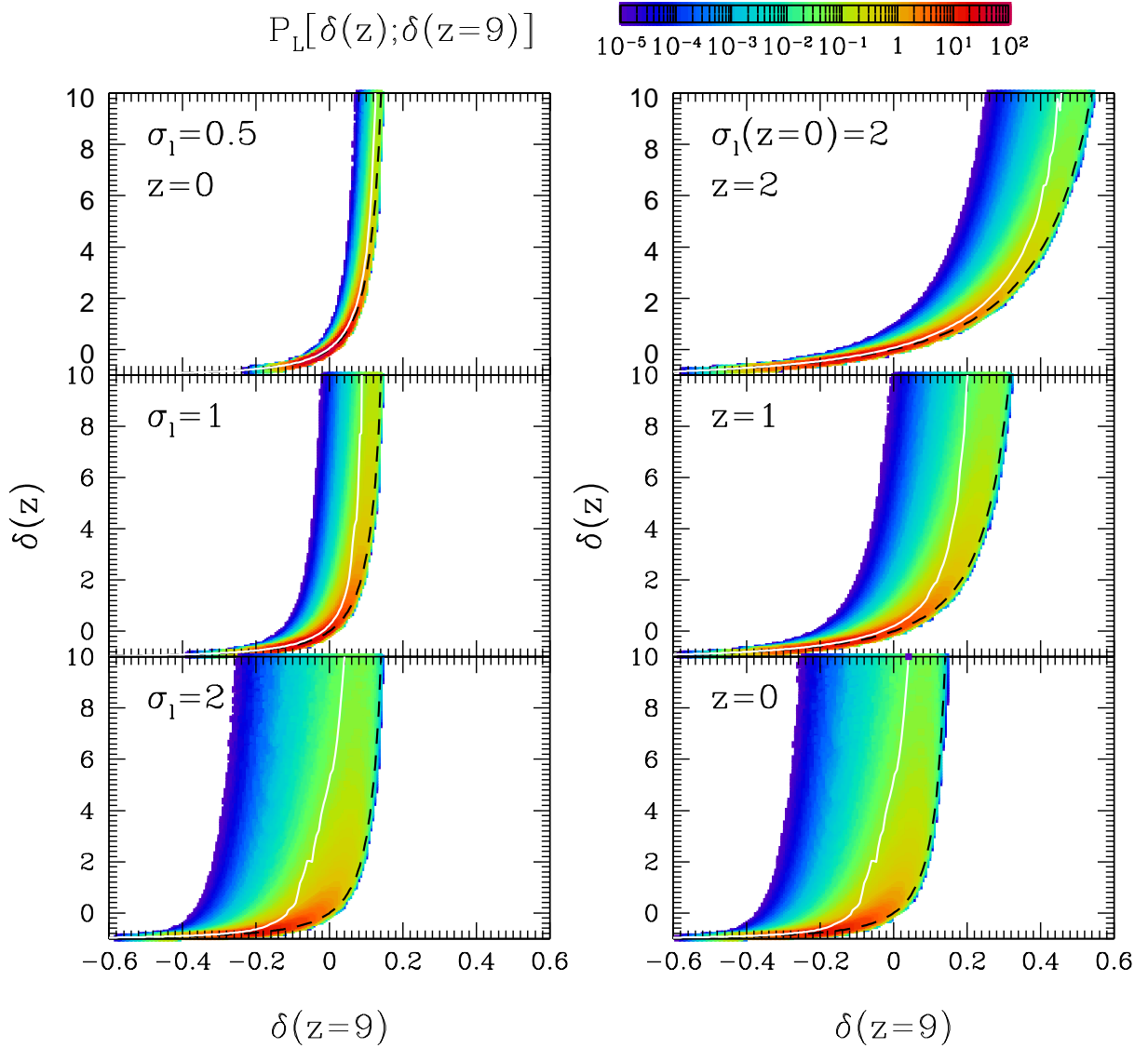}
 \figcaption{Contour plot of Lagrangian joint PDF
  $\probL(\delta(z),\delta(z=9))$ obtained from the ellipsoidal collapse
  model with linear external tide. 
  The left panel shows the results fixing the redshift to $z=0$, while the
  right panel represents the evolution of joint PDF fixing the linear
  variance to $\sigma_l=2$. 
  In each panel, solid lines indicate the conditional mean
  $[\delta(z)]_{\delta(z=9)}$ computed from joint PDF. 
  For comparison, the one-to-one local mapping obtained from the spherical
  collapse model are also plotted in short-dashed lines. 
  {\it Left}: $\sigma_l=0.5$({\it top}), $\sigma_l=1$({\it middle}) and
  $\sigma_l=2$({\it bottom}). 
  {\it Right}: $z=2$({\it top}), $z=1$({\it middle}) and $z=0$({\it bottom}). 
 \label{fig:jointPDF} }
\end{figure}

%
%
%
%
%
%
%
%
Next we focus on the joint PDFs. 
Left panel of Figure \ref{fig:jointPDF} shows the Lagrangian joint PDF
$\probL(\delta(z=0);\delta(z=9))$ from the ellipsoidal collapse model,
evaluated at the present time $z=0$ with various linear variances
$\sigma_l$. 
On the other hand, right panel of Figure \ref{fig:jointPDF} represents
the results fixing the linear fluctuation value to $\sigma_l=2$ at
present, but at different output times. 
Although Figure \ref{fig:jointPDF} does not rigorously correspond to the
N-body results obtained by \citet{Ka2001} (c.f. Fig.7 in their paper),
the qualitatively similar features can be drawn in several points. 
First, the scatter between $\delta(z=0)$ and $\delta(z=9)$ becomes
broader as increasing the time and the linear variance 
({\it top} to {\it bottom} in left panel). 
Second, the nonlinearity between the initial and the evolved density
indicated from the curvature of the conditional mean
$[\delta(z)]_{\delta(z=9)}$ ({\it solid}) also tends to increase as time
elapses ({\it top} to {\it bottom} in right panel). 
The one-to-one mapping obtained from the spherical collapse model 
({\it short-dashed}) is very different from the mean relation
$[\delta(z)]_{\delta(z=9)}$, but their mean relations basically 
reflect the qualitative behavior of local dynamics in 
Figure \ref{fig:evolve}.  
That is, the evolved results of local density in the ellipsoidal collapse 
tends to take a larger value than that in the spherical collapse. 
Moreover, recall the fact that both the initial and the final PDFs of 
local density $\probE(\delta)$ show a good agreement between the 
spherical and the ellipsoidal collapse model (see Fig.\ref{fig:1ptEulPDF}). 
This is indeed the same situation as in the N-body simulation; apart
from the detailed differences, a simple model of PDFs provides an
essential ingredient for the stochastic nature of N-body results. 
In this sense, the local approximation with ellipsoidal collapse models
can be regarded as a consistent and physical model of one-point 
statistics, which successfully explains the N-body simulations.

\begin{figure}
 \plotone{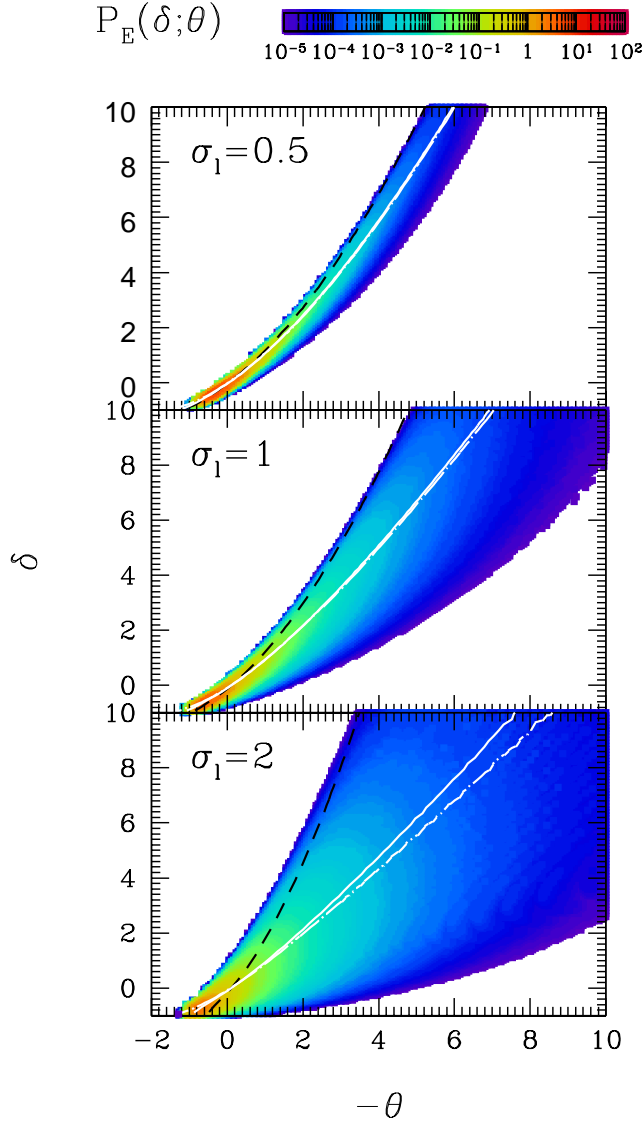}
 \figcaption{Contour plot of Eulerian joint PDF $\probE(\delta;\theta)$ 
   as functions of $-\theta$ and $\delta$ from the ellipsoidal 
   collapse model with linear external tide. In each panel, 
   the solid and the dot-dashed lines respectively denote the conditional 
   means $-[ \theta]_{\delta}$ and $[\delta]_{\theta}$. 
   The short-dashed lines represent the relation obtained from the 
   spherical collapse model: $\sigma_l=0.5$({\it top}); 
$\sigma_l=1$({\it middle});  $\sigma_l=2$({\it bottom}).
 \label{fig:thetadelta}}
\end{figure}

Finally, using the ellipsoidal collapse model with linear external tide,
we examine the Eulerian joint PDF of local density and
velocity-divergence evaluated at the same time, i.e.,
$\probE(\delta;\theta)$. 
In Figure \ref{fig:thetadelta}, contour plots of joint PDF
$\probE(\delta;\theta)$ for various linear fluctuation values $\sigma_l$
are depicted as function of $-\theta$ and $\delta$. 
This is the so-called {\it density-velocity relation}, which might be of
observational interest in measuring the density parameter $\Omega_{\rm
m}$ from the velocity-density comparison through the POTENT analysis
\citep[e.g.,][]{BD1989}. 
Along the line of this, theoretical works based on the Eulerian
perturbation theory have attracted much attention, as well as the N-body
study \citep[e.g.,][]{B1992a,CL1997,BCLSK1999}. 
Based on the solution of the local approximation (\ref{PDFdeltatheta}),
one can easily calculate the perturbation series of velocity-density
relation $[\theta]_{\delta}$ as function of local density and
density-velocity relation  $[\delta]_{\theta}$ as function of
velocity-divergence, the leading-order results of which are expected to
coincide with previous early works in the non-smoothing case, exactly. 
Beyond the perturbation analysis, Figure \ref{fig:thetadelta} reveals
the general trend of the stochastic nature in the velocity-density
relation. 
As increasing $\sigma_l$, the scatter becomes much broader and the
conditional means $[\delta]_{\theta}$ ({\it dot-dashed}) and
$[\theta]_{\delta}$({\it solid}) does not coincide with each other. 
Of course, the one-to-one mapping obtained from the spherical collapse
model ({\it short-dashed}) fails to match the both conditional means. 
These qualitative behavior is in fact consistent with the N-body results
by \citet{BCLSK1999} and the present model provides a simple way to
derive the non-linear and stochastic velocity-density relation. 
%
%
%
%
%
%
%
%
%
%
%
%
%
%
%
%
\section{CONCLUSION AND DISCUSSION}
\label{sec: conclusion}
%
%
%
%
%
%
%
%
%
%
%
%
In this paper, starting from a general theory of evolution of one-point
PDFs, we derived the evolution equations for PDF and within a local
approximation, consistent formal solutions of PDF are constructed in
both the Lagrangian and the Eulerian frames 
(see eqs.[\ref{LagPDFdelta}][\ref{LagPDFtheta}] for Lagrangian PDFs and 
eqs.[\ref{EulPDFdelta}][\ref{EulPDFtheta}] with 
eqs.[\ref{eq: function_g}][\ref{eq: function_h}] for Eulerian PDFs). 
In order to reveal the stochastic nature arising from the multivariate
Lagrangian dynamics, we further consider the Eulerian joint PDF
$\probE(\delta,\theta;t)$ (eq.[\ref{PDFdeltatheta}]) and the Lagrangian
joint PDF $\probL(\delta,t;\delta',t')$ (eq.[\ref{jointpdf}]). 
Then, adopting the spherical and the ellipsoidal collapse models as
representative Lagrangian local dynamics, we explicitly evaluate the
Eulerian PDFs, $\probE(\delta)$ and $\probE(\theta)$, as well as the
joint PDFs. 
The results from the ellipsoidal collapse model show several distinct
properties. 
While the PDF $\probE(\delta)$ almost coincides with the one-to-one
mapping of the spherical collapse model, the tails of
velocity-divergence PDF $\probE(\theta)$ largely deviate from those
obtained from the spherical model. 
These behaviors have also been confirmed from the perturbative analysis
of higher order moments. 
On the other hand, evaluating the Lagrangian joint PDF of local density,
$\probL(\delta,t;\delta',t')$, a large scatter in the relation between
the initial and the evolved density fields was found and their mean
relation fails to match the one-to-one mapping of spherical collapse
model. 
This remarkably reproduces the same situation in the N-body simulation. 
Therefore, the local approximation with ellipsoidal collapse model
provides a simple and physically reasonable model of one-point
statistics, consistent with the leading-order results of {\it exact}
perturbation theory.

Since the present formalism described in section \ref{sec: evolve_PDF}
is quite general, the approach does not restrict its applicability to
the pressure-less cosmological fluid. 
Rather, one may apply to the various fluid systems in presence or
absence of gravity.  
As mentioned in section \ref{subsec: LocalApprox}, however, the
applicability or the validity of local approximation of PDFs, in
principle, sensitively depends on the choice of Lagrangian local
dynamics. 
In the last section, simple and intuitive examples were examined for the
illustrative purposes. 
The results indicate that the multivariate Lagrangian dynamics rather
than the local model with a single variable can describe various
statistical features of fluid evolution including the stochastic nature.

Perhaps, a straightforward extension of the present treatment is to
include the effect of redshift-space distortion or projection effect,
which is practically important for proper comparison with observation. 
Before addressing this issue, however, remember the most primarily
importance of the smoothing effect. 
While the appropriate prescription for top-hat smoothing filter does
exist in the local approximation with the spherical collapse model
\citep[e.g.,][]{B1994b,PS1997,FG1998a}, the smoothing effect on the
approximation using ellipsoidal collapse model still needs to be
investigated. 
This step is in particular a crucial task in order to construct a more
physical prescription for one-point statistics of cosmic fields  and the
analysis is now in progress. 
The results will be presented elsewhere 
(Ohta, Kayo \& Taruya, in preparation). 
\acknowledgments

We are grateful to Y.Suto for reading the manuscript and comments. 
I.K acknowledges the support by Takenaka-Ikueikai Fellowship. 
This work is supported in part by the grand-in-aid for Scientific
Research of Japan Society for Promotion of Science (No.$1470157$). 
\clearpage 
\appendix
\section{ON NORMALIZATION CONDITION OF PDFs}
\label{appen: calculation_of_PDFs}
%
%
%
In computing the PDFs from the local approximation with the spherical or
the ellipsoidal collapse model, one may practically encounter the case
when the evolved density at a finite elapsed time infinitely diverges
for some regions in initial parameter space. 
To avoid the unphysical divergences, we restrict the integral over the
entire initial parameter space appearing in the PDFs to some regions
$V(t)$, in which the local density $\delta$ does not diverge. 
This modification slightly alters the normalization condition for both
the Lagrangian and the Eulerian PDFs in section 
\ref{subsec: LocalApprox} and \ref{subsec: Joint PDF}, which should be
corrected in the following re-normalization procedure.

First of all, the initial distribution function $\probI(\bfp)$ should be
re-normalized as 
\begin{equation}
\tilde{\probI}(\bfp) =
\frac{\displaystyle \probI(\bfp)}{\displaystyle 
\int_{V(t)}\prod_idp_i\,\probI(\bfp)}.
\label{eq: renorm_probI}
\end{equation}
Then, the modification of Lagrangian PDFs is to replace 
the initial distribution $\probI(\bfp)$ with $\tilde{\probI}(\bfp)$, 
together with the change of the integral region. 
For example, the one-pint PDF $\probL(\delta)$ in equation
(\ref{LagPDFdelta}) is modified as follows: 
\begin{equation}
 \probL (\delta;t)=\int_{V(t)}\prod_idp_i \tilde{\probI}(\bfp)\,\,
\delta_D(\delta-f(\bfp,t)).
\end{equation}
On the other hand, for the Eulerian PDFs, notice the fact that the
re-normalization (\ref{eq: renorm_probI}) also affects the relation
between $\delta$ and $f(\bfp,t)$ (see eq.[\ref{eq: function_g}]), which
must be modified as 
\begin{equation}
 1+\tilde{g}(\bfp,t)\equiv \tilde{N}_E(t)\left\{1+f(\bfp,t)\right\}
~~;~~~~~~~ \tilde{N}_E(t)\equiv \int_{V(t)}\prod_i dp_i\,\,
\frac{\tilde{\probI}(\bfp)}{1+f(\bfp,t)}.
\end{equation}
Using this relation, the renormalized one-point PDFs (\ref{EulPDFdelta})
and (\ref{EulPDFtheta}) respectively become 
\begin{equation}
 \probE(\delta;t)=\frac{1}{1+\delta}\,\int_{V(t)} \prod_i dp_i\,\, 
\tilde{\probI}(\bfp)\,\, \delta_D(\delta-\tilde{g}(\bfp,t))
\end{equation}
for the PDF $\probE(\delta)$ and  
\begin{equation}
 \probE(\theta)=\int_{V(t)} \prod_i dp_i\,\,
\frac{\tilde{\probI}(\bfp)}{1+\tilde g}\,\,\delta_D\left(\theta+\frac{1}{H
(1+f)}\frac{d f}{d t}+k\right)
\end{equation}
for the PDF $\probE(\theta)$. 
Here the variable $k$ is given by 
\begin{equation}
 k=-\frac{1}{\tilde{N}_E}\int_{V(t)} \prod_i dp_i\,\,
\frac{1}{H(1+f)^2}\,\,\frac{d f}{d t}\,\,\tilde{\probI}(\bfp),
\end{equation}
under the approximation that the time evolution of $V$ is neglected,
$\dot{V}/V\ll \dot{f}/(1+f)$.
Similarly, the renormalized Eulerian joint PDF $P(\delta,\theta;t)$ 
corresponding to the expression (\ref{PDFdeltatheta}) becomes 
\begin{equation}
 \probE(\delta,\theta;t)=\frac{1}{1+\delta}
\int_{V(t)} \prod_i dp_i\,\, \tilde{\probI}(\bfp)\,\,\delta_D(\delta-\tilde{g})\,
\delta_D\left(\theta+\frac{1}{H(1+f)}\frac{d f}{d t}+k\right).
\end{equation}
%
%
%
%
%
%
%
%
%
%
\section{CUMULANTS FROM SPHERICAL AND ELLIPSOIDAL COLLAPSE MODELS}
\label{appen: perturbation}
%
%
%
%
%
%
%
%
%
In this appendix, based on the local approximation with spherical and
ellipsoidal collapse models, we briefly summarize the essence of the
perturbation analysis for higher order moments of local density and
velocity-divergence. 
The details of the calculation procedure including the effect of
smoothing will be presented elsewhere 
(Ohta, Kayo \& Taruya, in preparation).  
Here, we only present the results in non-smoothing case.

Let us write down the evolution equation for ellipsoidal collapse model. 
In an Einstein-de Sitter universe, equation (\ref{EllipAxis}) becomes 
\begin{equation}
 a^2\frac{d^2\alpha_i}{da^2}-\frac{a}{2}\frac{d\alpha_i}{da}=
-\frac{3}{2}\alpha_i\left(\frac{1+\delta}{3}+\frac{b_i}{2}\delta 
+\lambda_{{\rm ext},i}\right). 
\label{eq: evolve_ellipsoid}
\end{equation}
In this case, the linear growth rate $D$ is proportional to the scale
factor $a$. 
Thus, the half length of principal axis $\alpha_i$ can be expanded by a
power series of $a$: 
\begin{equation}
 \alpha_i=a\left(1-\sum_{j=1}\gamma_i^{(j)}a^j\right)
\label{eq: expanded_alpha}
\end{equation}
Note that the initial conditions (\ref{eq: init_con(1)}) and 
(\ref{eq: init_con(2)}) state $\gamma_i^{(1)}=\lambda_i$. 
Hence, substituting the expression (\ref{eq: expanded_alpha}) into 
(\ref{eq: evolve_ellipsoid}) and solving the evolution equation
perturbatively, the coefficient $\gamma_i^{(j)}$ is systematically
determined order by order and is expressed as the function of
$\lambda_i$. 
Thus, the perturbative expansion of local density $\delta$ given by
(\ref{eq: delta_ellipsoid}) is also expressed in terms of  $\lambda_i$ : 
\begin{equation}
 f(\lambda_i,a)=\frac{a^3}{\alpha_1\alpha_2\alpha_3}-1
=\sum_{j=1}\, \delta^{(j)}(\lambda_i)\,\,\,a^j, 
\end{equation}
with the corresponding boundary condition being 
$\delta^{(1)}=\lambda_1+\lambda_2+\lambda_3$.

Provided the function $f(\lambda_i,a)$, one can calculate the $n$-th
order moments of $\delta$ and $\theta$ as well as the normalization
factor $N_E$, according to the one-point PDFs of local approximation,
(\ref{EulPDFdelta}) and (\ref{EulPDFtheta}):   
\begin{eqnarray}
&N_E = & \int \prod_i\,d\lambda_i\,\,\frac{\probI(\lambda_i)}{1+f(\lambda_i)}, 
\label{eq: N_E_ellipsoid}
\\
 &\langle\delta^n\rangleE =& \int d\delta\ \delta^n\,\,\probE(\delta)
=\int \prod_id\lambda_i\,\,
\frac{[N_E\{1+f(\lambda_i,a)\}-1]^n}{N_E\{1+f(\lambda_i)\}}
\probI(\lambda_i),
\label{eq: moment_delta}
\\
\nonumber
\\
& \langle\theta^n\rangleE =& \int d\theta\ \theta^n\,\,\probE(\theta)
=\int \prod_id\lambda_i\,\,
\frac{1}{N_E(1+f)}\left(-\frac{a}{1+f}\frac{d f}{d a}
-\frac{a}{N_E}\frac{d N_E}{d a}\right)^n\probI(\lambda_i),
\label{eq: moment_theta}
\end{eqnarray}
where the function $\probI(\lambda_i)$ denotes the PDF of initial
parameter $\lambda_i$ given by (\ref{DoroPDF}).

Below, we separately present the perturbation results in each model. 
For the results of higher order moments, i.e., variance, skewness and 
kurtosis given by (\ref{eq: perturb_sigma})--(\ref{eq: perturb_T4}), 
numerical values of the perturbation coefficient are summarized in table
\ref{tbl : loop_correction} and the departure from the leading-order
results are depicted in figure \ref{fig:MomentDelta}. 
%
%
%
%
\subsection{Ellipsoidal collapse model with linear external tide}
%
%
%
%
%
%
If adopting the ellipsoidal collapse model with linear external tide
(see eq.[\ref{eq: external_tide}]), the perturbative expansion of local
density, $\delta^{(n)}$ up to the fifth order becomes 
\begin{eqnarray}
 \delta^{(1)}&=&\delta_l,\\
\delta^{(2)}&=& \frac{17}{21}\delta_l^2+\frac{4}{21}J_1,\\
\delta^{(3)}&=&\frac{341}{567}\delta_l^3+\frac{1538}{4725}\delta_lJ_1+\frac{4}{405}J_2,\\
\delta^{(4)}&=&\frac{55805}{130977}\delta_l^4+\frac{952144}{2480625}\delta_l^2J_1+\frac{345088}
{16372125}\delta_lJ_2+\frac{12368}{363825}J_1^2,\\
\delta^{(5)}&=&\frac{213662}{729729}\delta_l^5+\frac{237342074}{621928125}\delta_l^3J_1+\frac{93363344}
{3192564375}\delta_l^2J_2\nonumber \\
 & &\ +\frac{52865818}{638512875}\delta_lJ_1^2+\frac{135052}{34827975}J_1J_2,
\end{eqnarray}
where $\delta_l$ denotes the linear fluctuation density given by 
$\delta_l=\lambda_1+\lambda_2+\lambda_3$. 
Here we introduced the quantities 
$J_1\equiv x^2+xy+y^2$ and $J_2\equiv (x-y)(2x+y)(x+2y)$  
with the variables $x$ and $y$ being 
$x=\lambda_1-\lambda_2$ and $y=\lambda_2-\lambda_3$, respectively.  
Substituting the above expansion results into (\ref{eq: N_E_ellipsoid}),  
the normalization factor of Eulerian PDF becomes 
\begin{equation}
 N_E=1+\frac{69668}{3898125}\sigma_l^4 + \cdots.
\end{equation}
Further, using the equations (\ref{eq: moment_delta}) and 
(\ref{eq: moment_theta}), one can obtain the variance, the skewness and
the kurtosis defined in 
(\ref{eq: perturb_sigma})--(\ref{eq: perturb_T4}). 
For the variances of local density and velocity-divergence, the
perturbative correction up to the two-loop order ${\cal O}(\sigma_l^6)$
becomes 
\begin{eqnarray}
 \sigma^2&=&\sigma_l^2+\frac{57137}{33075}\sigma_l^4+
\frac{469828713881}{111739753125}\sigma_l^6+\cdots,
\label{elllindsigma}
\\
 \sigma_\theta^2 &=& \sigma_l^2+\frac{8747}{11025}\sigma_l^4+
\frac{154583563}{165540375}\sigma_l^6+\cdots. 
\label{elllintsigma}
\end{eqnarray}
As for the skewness and the kurtosis, we obtain the results up to the 
one-loop order ${\cal O}(\sigma_l^2)$:  
\begin{eqnarray}
 S_3&=&\frac{34}{7}+\frac{646404856}{63669375}\sigma_l^2+\cdots, 
\label{elllinS3}
\\
 S_4&=&\frac{60712}{1323}+\frac{210688932175742}{782178271875}\sigma_l^2
+\cdots,
\label{elllinS4}
\end{eqnarray}
for the local density and 
\begin{eqnarray}
 T_3 &=& -\frac{26}{7}-\frac{333940984}{63669375}\sigma_l^2+\cdots,
\label{elllinT3}
\\
 T_4 &=& \frac{12088}{441}+\frac{17145801103334}{156435654375}\sigma_l^2
+\cdots,
\label{elllinT4}
\end{eqnarray}
for the velocity-divergence. 
%
%
%
%
\subsection{Ellipsoidal collapse model with non-linear external tide}
%
%
%
%
%
%
In the case of the ellipsoidal collapse model with nonlinear external
tide, the perturbative solution $\delta^{(n)}$ up to the fifth order
becomes 
\begin{eqnarray}
 \delta^{(1)}&=&\delta_l,\\
 \delta^{(2)}&=&\frac{17}{21}\delta_l^2+\frac{4}{21}J_1,\\
 \delta^{(3)}&=&\frac{341}{567}\delta_l^3+\frac{338}{945}\delta_lJ_1+\frac{92}{3969}J_2,\\
 \delta^{(4)}&=&\frac{55805}{130977}\delta_l^4+\frac{485288}{1091475}\delta_l^2J_1+\frac{234088}
{4584195}\delta_lJ_2+\frac{429728}{10696455}J_1^2,\\
 \delta^{(5)}&=&\frac{213662}{729729}\delta_l^5+\frac{292398464}{638512875}\delta_l^3J_1+\frac{64182728}
{893918025}\delta_l^2J_2\nonumber \\
 & &\ +\frac{6541246}{59594535}\delta_lJ_1^2+\frac{828974992}{96364363095}J_1J_2.
\end{eqnarray}
Then the normalization factor is 
\begin{equation}
 N_E= 1+\frac{10844}{848925}\sigma_l^4+\cdots.
\end{equation}
The variances of local density and velocity-divergence up to the
two-loop order become 
\begin{eqnarray}
 \sigma^2&=&\sigma_l^2+\frac{439}{245}\sigma_l^4+\frac{3143785639}{695269575}\sigma_l^6+\cdots,
\label{ellnoldsigma}
\\
 \sigma_\theta^2&=&\sigma_l^2+\frac{145}{147}\sigma_l^4+\frac{1708470649}{1158782625}\sigma_l^6 + \cdots.
\label{ellnoltsigma}
\end{eqnarray}
The results for the skewness and the kurtosis can be reduced to 
\begin{eqnarray}
 S_3&=&\frac{34}{7}+\frac{1041064}{101871}\sigma_l^2+\cdots, 
\label{ellnolS3}
\\
 S_4&=&\frac{60712}{1323}+\frac{941370178286}{3476347875}\sigma_l^2+\cdots, 
\label{ellnolS4}
\end{eqnarray}
for the local density and 
\begin{eqnarray}
 T_3 &=& -\frac{26}{7}-\frac{2701112}{509355}\sigma_l^2+\cdots,
\label{ellnolT3}
\\
T_4 &=& \frac{12088}{441}+\frac{128186956538}{1158782625}\sigma_l^2 +\cdots.
\label{ellnolT4}
\end{eqnarray}
for the velocity-divergence.
%
%
%
%
\subsection{Spherical collapse model}
%
%
%
%
%
%
In the case of spherical collapse model, the perturbative solutions of
local density $f(\delta_l)$ just corresponds to those obtained from the
ellipsoidal collapse model simply setting $J_1=0$ and $J_2=0$, since the
three initial parameters of principal axis $\lambda_i$ are identical. 
The cumulants are then calculated in similar way to the previous
subsection except for the initial parameter PDF (\ref{eq: gauss_PDF}). 
The resultant normalization factor is 
\begin{equation}
 N_E=1+\frac{4}{21}\sigma_l^2+\frac{460}{43659}\sigma_l^4+\cdots. 
\end{equation}
The variances of $\delta$ and $\theta$ become 
\begin{eqnarray}
 \sigma^2&=& \sigma_l^2+\frac{1909}{1323}\sigma_l^4+
\frac{344439415}{107270163}\sigma_l^6+\cdots,
\\
 \sigma_\theta^2&=&\sigma_l^2-\frac{11}{147}\sigma_l^4+
\frac{319159}{1324323}\sigma_l^6+\cdots.
\end{eqnarray}
The skewness and the kurtosis respectively becomes 
\begin{eqnarray}
 S_3&=&\frac{34}{7}+\frac{1026488}{101871}\sigma_l^2+\cdots,
\\
 S_4&=&\frac{60712}{1323}+\frac{22336534498}{83432349}\sigma_l^2+\cdots,
\end{eqnarray}
for the local density and 
\begin{eqnarray}
 T_3&=&-\frac{26}{7}-\frac{613936}{101871}\sigma_l^2+\cdots,
\\
 T_4&=&\frac{12088}{441}+\frac{10081115810}{83432349}\sigma_l^2+\cdots,
\end{eqnarray}
for the velocity-divergence.  
\clearpage
%
%
%
%

%
%
%
%
%
%
\clearpage
%
%
%
%
%
\begin{deluxetable}{ccccccc}
\tablecolumns{4}
\tablewidth{9cm}
\tablecaption{Coefficients of perturbative correction for the density
 and the velocity-divergence in one-point statistics from the spherical
 and the ellipsoidal collapse models (see 
 eqs.[\ref{eq: perturb_sigma}]--[\ref{eq: perturb_T4}]).  
\label{tbl : loop_correction}}
\tablehead{
\colhead{~~} &~~& \colhead{SCM$^{~\dag}$} &~~&  \colhead{ECM1$^{~\ddag1}$}
&~~& \colhead{ECM2$^{~\ddag2}$}
}
\startdata
$s_{2,4}$ &~~& $1.44$~~~~ &~~& $1.73$~~~~ &~~& $1.79$~~  \\
$s_{2,6}$ &~~& $3.21$~~~~ &~~& $4.20$~~~~ &~~& $4.52$~~    \\
\hline
$S_{3,0}$ &~~& $4.86$~~~~ &~~& $4.86$~~~~ &~~& $4.86$~~  \\
$S_{3,2}$ &~~& $10.08$~~~~ &~~& $10.15$~~~~ &~~& $10.22$~~ \\
\hline
$S_{4,0}$ &~~& $45.89$~~~~ &~~& $45.89$~~~~ &~~& $45.89$~~  \\
$S_{4,2}$ &~~& $267.72$~~~~ &~~& $269.36$~~~~ &~~& $270.79$~~  \\
\hline\hline
$s_{2,4}^{\theta}$ &~~& $-0.075$~~~~ &~~& $0.79$~~~~ &~~& $0.99$~~  \\
$s_{2,6}^{\theta}$ &~~& $0.24$~~~~ &~~& $0.93$~~~~ &~~& $1.47$~~   \\
\hline
$T_{3,0}$ &~~& $-3.71$~~~~ &~~& $-3.71$~~~~ &~~& $-3.71$~~  \\
$T_{3,2}$ &~~& $-6.03$~~~~ &~~& $-5.24$~~~~ &~~& $-5.30$~~  \\
\hline
$T_{4,0}$ &~~& $27.41$~~~~ &~~& $27.41$~~~~ &~~& $27.41$~~  \\
$T_{4,2}$ &~~& $120.83$~~~~ &~~& $109.60$~~~~ &~~& $110.62$~~ \\
\enddata
\vspace{12pt}\par\noindent
$^{\dag}$ Spherical collapse model\\
$^{\ddag1}$ Ellipsoidal collapse model with linear external tide\\
$^{\ddag2}$ Ellipsoidal collapse model with non-linear external tide
\end{deluxetable}
\clearpage
\end{document}